\documentclass{ws-ijmpa}
\usepackage{xcolor}
\usepackage[sort,compress]{cite}
\usepackage[verbose]{hyperref}
\hypersetup{colorlinks=false,allbordercolors=blue,pdfborderstyle={/S/U/W 1}}
\usepackage{graphicx}
\usepackage{multirow}
\usepackage{amsmath,amssymb,amsfonts}
\usepackage{tabularx}
\usepackage[title]{appendix}
\usepackage{xcolor}
\usepackage{textcomp}
\usepackage{newunicodechar}
\usepackage{manyfoot}
\usepackage{breqn}
\usepackage{booktabs}
\usepackage{algorithm}
\usepackage{algorithmicx}
\usepackage{algpseudocode}
\usepackage{listings}
\usepackage{float}
\usepackage{hyperref}
\usepackage{mathtools}
\usepackage{fix-cm}
\usepackage{rotating}
\usepackage{float}
\usepackage{subcaption}
\usepackage{enumerate}
\usepackage{dirtytalk}
\usepackage{microtype}
\usepackage{siunitx}
\usepackage[utf8]{inputenc}
\usepackage{wrapfig}
\usepackage{bm}
\usepackage{array}
\usepackage{textcomp}
\usepackage{paralist}
\begin{document}

\markboth{Pradosh Keshav MV, Jithesh V, Kenath Arun}{}

\catchline{}{}{}{}{}

\title{Quintessence–Chameleon Transitions in Anisotropic Kiselev Model of Neutron Stars}

\author{Pradosh Keshav MV\(^{1}\)} \author{V Jithesh*}  \author{Kenath Arun*\(\dagger\) }
\address{\(^{1*}\)Department of Physics and Electronics, Christ University\\
Bangalore, Karnataka 560029,
India\\
\(^{\dagger}\)kenath.arun@christuniversity.in}

\maketitle

\begin{abstract}
We investigate a chameleon scalar field dynamically interacting with a Kiselev-type metric, where the static anisotropic fluid part of the metric is replaced by a density-dependent scalar field non-minimally coupled to curvature. This construction enables a transition from screened behavior in high-density regions—where the scalar acquires an effective mass $m_\phi \propto \rho^{1/2}$—to unscreened quintessence dynamics at large scales, characterized by a critical screening radius $r_{\rm crit} \propto m_\phi^{-1}$. By solving the modified Tolman-Oppenheimer-Volkoff equations under spherical symmetry, we show that radial scalar gradients $(\partial_r \phi)$ induce pressure anisotropies $(\Delta p \propto r^{-1})$ in neutron star envelopes, while deviations from general relativity (GR) are suppressed deep in the core $(r < r_{\rm crit} \sim 0.03\,\mathrm{km})$ without destabilizing it. We further demonstrate that increasing the scalar coupling $\beta$ enhances scalar energy density, which counteracts anisotropic pressure support and slightly reduces the maximum mass. The resulting (stable) configurations yield maximum mass  $M_{\text{max}} \approx 1.75 \pm 0.280\,M_\odot$ and radii $R \approx 11.35 \pm 1.11 \,\mathrm{km}$, consistent with conservative upper and lower bounds from multimessenger observations. Scalar contributions induce a modest suppression in the dimensionless tidal deformability $(\Lambda_{\rm ST}/\Lambda_{\rm GR} \sim 0.137)$ at $1.4\,M_\odot$ i.e, 86\% suppression compared to typical GR star, falling well within the LIGO/Virgo range $70 \lesssim \Lambda_{1.4} \lesssim 580$. These results demonstrate that environmentally screened scalar fields can dynamically generate anisotropies and modify neutron star structure without violating current astrophysical bounds. In particular, percent-level deviations in compactness and tidal response provide falsifiable signatures of short-range scalar forces, offering a novel target for next-generation gravitational wave and multimessenger surveys.

\end{abstract}

\section{Introduction}
Observations of the accelerated expansion of the universe have established the existence of a pervasive, unknown energy field—dark energy—characterized by a negative equation of state (EoS) \( w \approx -1 \) \cite{perlmutter1999measurements, riess1998observational}. While the \(\Lambda\)-Cold Dark Matter (\(\Lambda\)-CDM) paradigm remains consistent with large-scale observations, including Cosmic Microwave Background (CMB) anisotropies \cite{planck2020cosmological}, baryon acoustic oscillations \cite{Yang2020Dynamical}, and weak gravitational lensing \cite{Huterer2001Weak}, its reliance on a static cosmological constant \(\Lambda\) faces unresolved theoretical challenges \cite{arun2018alternate}. Furthermore, tensions at galactic scales—such as discrepancies in halo mass profiles, satellite galaxy counts \cite{del2017small}, and the \(H_0\) tension—suggest either missing astrophysical processes or fundamental limitations of \(\Lambda\)-CDM (see \cite{mavromatos2022lambda} for a discussion at small galactic scales). Dynamical scalar field models, such as quintessence \cite{peebles2003cosmological}, K-essence \cite{armendariz2001essentials}, and chameleon theories \cite{khoury2004chameleon}, address these issues by introducing spacetime-dependent scalar degrees of freedom. The chameleon field, in particular, evades solar system constraints through a density-dependent effective mass, enabling it to mimic dark energy’s EoS at cosmological scales while remaining screened in high-density regions \cite{khoury2013chameleon}. This screening mechanism distinguishes it from other scalar models, which often require fine-tuned potentials to avoid fifth-force detections \cite{Dima2021Dynamical}. While dark energy’s cosmological effects are well-studied, its behavior in extreme curvature regimes—such as near neutron stars and black holes—remains poorly constrained \cite{brax2017neutron, li2016neutron}. Scalar fields, if present in these environments, would generate anisotropic stresses, crucial for examining potential deviations from General Relativity (GR) in strong-field regimes with signatures detectable through gravitational waves and electromagnetic probes \cite{Vlachos2021Echoes, Qin2022Polarized, Cañate2021Gravitational}.

Recent studies \cite{DeCesare2022Evolving, Davis2014Astrophysical} have shown that scalar fields stabilize black holes near trapping horizons through metric parametrizations sensitive to horizon displacement and expansion rates. In this context, it is quite natural to investigate the role of scalar fields in the processes of galaxy formation and the evolution of more compact astrophysical structures, such as stars and their clusters. Compact astrophysical objects—neutron stars, black holes, and exotic stars—serve as natural laboratories for probing scalar dynamics in extreme curvature regimes \cite{ottoni2024x}. Their strong gravitational fields and anisotropic matter distributions destabilize isotropic fluid approximations, necessitating relativistic treatments of pressure anisotropy (\(\Delta p \equiv p_t - p_r \neq 0\)). In such configurations, anisotropies could arise from nuclear matter shear stresses, magnetic field geometries, or non-minimal scalar-tensor couplings \cite{maurya2021anisotropic}. For example, scalarization—a phase transition where compact objects develop scalar hair—occurs in Damour-Esposito-Farèse scalar-tensor theory when the Tolman-Oppenheimer-Volkoff (TOV) equations admit bifurcation solutions beyond critical compactness \cite{barausse2013neutron}. Similarly, spin-induced scalarization in Kerr black holes \cite{Herdeiro2020Spin-Induced}, spontaneous scalarization in neutron stars \cite{Wong2019Effective}, and spontaneous vectorization in Einstein-Maxwell-scalar systems \cite{silva2022ghost} demonstrate how anisotropy-generating mechanisms depend on spacetime symmetry and coupling topology. These phenomena are influenced by non-minimal couplings between scalar fields and curvature, often requiring modifications to the gravitational action and effective stress-energy tensor.\cite{koivisto2015scalar}.

The Krori–Barua (KB) metric \cite{Krori1975} is a widely utilized tool for modeling compact star studies within general relativity as well as in modified gravity theories such as \(f(R)\), \(f(G)\), and \(f(T)\) \cite{abbas2014cylindrically, abbas2015strange, abbas2015anisotropicfG, zubair2016possible}. Although the KB metric has been used to investigate various aspects of these stars—including anisotropic strange stars under the assumption \(p = \alpha\rho\) (with \(0 < \alpha < 1\)) \cite{abbas2015anisotropic} that mimic quintessence dark energy—local impacts (such as the dynamic impact of scalar-curvature coupling) of the real scalar field remains problematic. These are the scalar fields that do not carry a charge, and constructing regular stable solutions is more difficult in this case. For instance, the known static solutions for real scalar fields tend to be either singular in nature \cite{wyman1981static, jetzer1992dynamical} or involve phantom scalar fields \cite{kodama1978general, kodama1979properties}. Singular solutions characterized by trivial topology for self-interacting fields, which involve ordinary scalar fields, are asymptotically flat; nevertheless, their stability remains uncertain. Unscreened scalar fields in such models predict tangential pressures (\(p_t > 10^{35} \, \text{dyne/cm}^2\)) incompatible with pulsar timing constraints \cite{antoniadis2013massive}, while regular solutions for real scalar fields often require unphysical potentials or violate energy conditions \cite{torii1999can}. Many studies (see \cite{nazar2023relativistic}) employ heuristic radial profiles to parameterize \(\Delta p\); however, these phenomenological treatments lack a solid foundation in first-principles scalar dynamics, especially in the absence of gravity or a perfect fluid approximation.

Existing models of anisotropy in compact objects often rely on phenomenological parameterizations of \(\Delta p(r)\) or neglect screening mechanisms necessary to suppress fifth forces in dense environments \cite{sakstein2013stellar, sakstein2014detecting, sakstein2015testing}. For instance, the Vaidya metric and Bowers-Liang formalism extend TOV solutions to include radial pressure anisotropy, predicting increased neutron star maximum masses \cite{bowers1974anisotropic}, but lack microphysical derivations of \(\Delta p\). Similarly, Horndeski theories \cite{galeev2021anisotropic, kennedy2018reconstructing} generate anisotropy via non-minimal kinetic couplings, while Einstein-aether theory imposes preferred-frame effects through timelike vector fields. Boson stars \cite{liebling2023dynamical}, gravastars \cite{mazur2004gravitational}, and fuzzy dark matter halos \cite{kulkarni2022if} replace singularities with anisotropic fluid or field configurations, often violating energy conditions (e.g., \(p_r < -\rho\)) for stability. These limitations highlight the need for an alternative mechanism that links scalar field dynamics to observable anisotropies without ad-hoc assumptions. Recent studies of black hole scalar hair in Einstein-scalar-Gauss-Bonnet gravity \cite{Herdeiro2020Spin-Induced} and neutron star mergers in Horndeski theories \cite{ventagli2025neutron, barausse2013neutron} further motivate our investigation into a real scalar field plus an anisotropic fluid even when the potential energy of the scalar field satisfies \(V(\phi)>0\). This mainly stems from the necessity for a self-consistent framework in which anisotropy emerges dynamically from scalar field gradients \((\partial_r \phi)\), providing a unified description of both non-relativistic stars and those where relativistic effects are significant, such as neutron stars.

In this paper, we address this gap by coupling a chameleon-like scalar field to the Kiselev metric \cite{kiselev2003quintessence}, a framework previously applied to static quintessence fluids \cite{heydarzade2017black}. Unlike prior studies, we replace the Kiselev metric’s homogeneous dark energy component with a chameleon field whose effective mass \(m_\phi(\rho)\) depends on local matter density. This dynamically suppresses fifth forces in high-density regions while amplifying anisotropies (\(\Delta p \propto (\partial_r \phi)^2\)) in low-density envelopes. Our formulation serves as an intermediate step for including the non-minimal interaction between the perfect fluid and the real scalar field that leads to regular solutions that are generally different from typical polytropic stars. Kiselev anisotropic fluid--quintessence as local impacts of cosmic acceleration upon the compact stars--has also been studied widely in the context of black hole shadows \cite{konoplya2019shadow, zeng2020shadows, abdujabbarov2017shadow}, quasinormal modes \cite{chen2005quasinormal, zhang2006quasinormal}, black holes thermodynamics \cite{thomas2012thermodynamics, saleh2018thermodynamics, toledo2019some, toledo2019reissner}, and modified gravity \cite{sakti2020kerr, heydarzade2017black, santos2023kiselev}. Additional properties of the Kiselev black hole and the collapse of null and timelike thin shells within this spacetime were considered in \cite{saadati2021thin, javed2024impact}. However, the presence of non-minimal interaction significantly alters both the radial matter distribution within the metric approximation of the star's interior and its total mass \cite{dzhunushaliev2011chameleon}. This is effectively captured for a scalar field whose mass grows as \(m_\phi \propto \rho^{1/2}\) in high-density environments (such as in neutron star cores), while its potential dominates in low-density regions (as in stellar envelopes or cosmological scales).

The Kiselev metric’s stress-energy tensor can be decomposed into a perfect fluid and an anisotropic field component (such as electromagnetic or scalar) \cite{Boonserm2019Decomposition}, enabling parameterization of deviations from perfect fluid behavior through a scalar field’s radial gradients. Unlike the Vaidya metric, which assumes outgoing radiation, or Bowers-Liang models reliant on ad-hoc \(\Delta p(r)\) ansätze, we generalize the Kiselev metric to include a density-dependent scalar field, allowing for dynamical anisotropies in compact objects. Taking a special case of neutron stars as prototypical compact objects, we numerically solve the modified TOV equations given by \cite{campitelli2024neutron}, assuming that the Kislev normalization factor \(a\) could be fixed using the scalar amplitude \(\phi_0\) and coupling strength \(\beta\). We analyze two critical constraints: the maximum mass of neutron stars and tidal deformability. The first constraint arises from NICER measurements of the most massive known pulsar, PSR J0740+6620, with mass $2.072^{+0.067}_{-0.066}\, M_\odot$ \cite{riley2021nicer}, thereby requiring that any viable EoS must support a maximum mass to be at least $2\, M_\odot$. The second constraint follows from the binary neutron star merger GW170817 \cite{Abbott2017GW170817:}, which places an upper bound on the dimensionless tidal deformability of a $1.4\, M_\odot$ star, typically denoted as $\Lambda_{1.4} \lesssim 580$ \cite{abbott2018gw170817} and radius \(R_{\rm 1.4} = 11.0^{+0.9}_{-0.6} \) km \cite{capano2020stringent}. However, such predictions often require multimessenger observables to look for compact objects that have a larger critical radius for a given mass than would be expected from spin-induced scalarization \cite{Herdeiro2020Spin-Induced}, where scalar hair emerges at larger radii independent of density.

The paper is organized as follows. We begin by establishing the Kiselev metric’s formalism, analyzing its anisotropic stress components and applicability to scalar field configurations in Section 2. Section 3 examines scalar field dynamics near compact objects, focusing on the effective potential and the conditions under which scalar fields behave like perfect fluids--linking its radial gradients to pressure anisotropies (\(\Delta p\)) and quintessence-like behavior. In Section 4, we analyze non-minimal couplings in the vicinity of compact objects, solving the scalar field equation of motion under quasi-static equilibrium and deriving the effective scalar mass \(m_\phi(\rho)\). Section 5 introduces the critical radius \(r_{\rm crit}\) demarcating screened and unscreened regimes, while Section 6 presents numerical solutions to the modified TOV equations, predicting neutron star mass-radius relations compared with GR. Section 7 discusses stability criteria and observational signatures of anisotropy, and Section 8 outlines future prospects for testing the model with multimessenger data.

Throughout sections 2–5, all derivations—including the formulation of the chameleon action and the scalar field equations—are performed in natural units (\(\hbar = c = 1\)). For our numerical analysis of neutron star structure in Section 6, however, we switch to geometric units by setting \(G = c = 1\). In these units, length, time, and mass are expressed in terms of kilometers (km) and solar masses (\(M_\odot\)), and curvature quantities (such as the Ricci scalar and scalar potential \(V(\phi)\)) have dimensions of inverse length squared (km\(^{-2}\)). When reporting astrophysical observables – for example, neutron star densities and pressures – we convert to CGS units (\(\text{g}/\text{cm}^{3}\) and \(\text{dyne}/\text{cm}^{2}\), respectively).

\section{Kiselev Metric and Approximation for Anisotropic Fluid Model}

The Kiselev metric \cite{kiselev2003quintessence} represents a spherically symmetric solution in general relativity that describes a black hole enveloped by an anisotropic fluid characterized by the EoS \(p = w_q \rho\). Although originally formulated on phenomenological grounds, this metric can be derived self-consistently within scalar-tensor theories of gravity by identifying the scalar field’s stress-energy tensor as the source corresponding to the anisotropic fluid. We begin with the general scalar-tensor action:
\begin{equation}
    S = \int d^4x \sqrt{-g} \left[ \frac{1}{16\pi} f(\phi) R - \frac{1}{2} \omega(\phi) (\nabla \phi)^2 - V(\phi) \right] + S_{\text{matter}}, \label{eq:gen action}
\end{equation}
where \(f(\phi)\) determines the non-minimal coupling between the scalar field \(\phi\) and the Ricci scalar \(R\), \(\omega(\phi)\) specifies the kinetic structure of \(\phi\), and \(V(\phi)\) is the scalar potential. Varying Eq.~\eqref{eq:gen action} yields the modified Einstein equations:
\begin{equation}
    f(\phi) G_{\mu\nu} + \nabla_\mu \nabla_\nu f(\phi) - g_{\mu\nu} \Box f(\phi) = 8\pi \left( T_{\mu\nu}^{(\phi)} + T_{\mu\nu}^{(\text{matter})} \right),\label{eq:mod eins}
\end{equation}
with the stress-energy tensor of the scalar field given by:
\begin{equation}
    T_{\mu\nu}^{(\phi)} = \omega(\phi) \nabla_\mu \phi \nabla_\nu \phi - \frac{1}{2} g_{\mu\nu} \left[ \omega(\phi) (\nabla \phi)^2 + 2V(\phi) \right].
\end{equation}

The Kiselev metric generalizes the Schwarzschild and Reissner-Nordström geometries through the parameterized EoS \(w_q\) \cite{chen2005quasinormal}. To recover this metric, we assume a static, spherically symmetric spacetime of the form:
\begin{equation}
    ds^2 = -f(r)dt^2 + f(r)^{-1}dr^2 + r^2 d\Omega^2, \label{eq:anstas kiselv}
\end{equation}
with the metric function:
\begin{equation}
    f(r) = 1 - \frac{2M}{r} - \frac{a}{r^{3w_q + 1}}, \label{eq:metric_function}
\end{equation}
where \(a\) is a normalization factor. The anisotropic fluid sourcing this metric is characterized by stress-energy components \(T^t_t = T^r_r = \rho\) and \(T^\theta_\theta = T^\varphi_\varphi = -\frac{1}{2}(3w_q + 1)\rho\). In order for the scalar field to mimic this structure, its stress-energy tensor must also exhibit similar anisotropic characteristics to the Kiselev metric. Considering a static scalar field \(\phi(r)\), the non-vanishing components of \(T^{(\phi)}_{\mu\nu}\) are:
\begin{align}
    {T^{(\phi)}_t}^t &= -\frac{1}{2} \left[ \omega(\phi) \left(\frac{d\phi}{dr}\right)^2 + 2V(\phi) \right], \label{eq:tphi1} \\
    {T^{(\phi)}_r}^r &= \frac{1}{2} \left[ \omega(\phi) \left(\frac{d\phi}{dr}\right)^2 - 2V(\phi) \right], \label{tphi2} \\
    {T^{(\phi)}_\theta}^\theta &= {T^{(\phi)}_\phi}^\phi = -\frac{1}{2} \left[ \omega(\phi) \left(\frac{d\phi}{dr}\right)^2 + 2V(\phi) \right]. \label{eq:emtensor}
\end{align}
Matching the condition \({T^{(\phi)}_t}^t = {T^{(\phi)}_r}^r\) immediately implies:
\begin{equation}
  \omega(\phi)\left(\frac{d\phi}{dr}\right)^2 = 0,
\end{equation}
which, for a minimally coupled field, forces either \(\omega(\phi)=0\) (yielding a potential-dominated scenario with \(w_q = -1\)) or \(\phi'(r)=0\) (resulting in a trivial scalar configuration). To obtain a non-trivial scalar field profile for a generic EoS \(w_q\), it is necessary to invoke a non-minimal coupling (\(f(\phi) \neq 1\)) or incorporate additional terms—such as Horndeski contributions of the form \(G^{\mu\nu} \nabla_\mu \phi \nabla_\nu \phi\)—which relaxes strict constraints and allow anisotropic stresses. By adopting a power-law potential \(V(\phi) \propto \phi^n\) and accommodating non-minimal couplings, one may solve the field equations to obtain a scalar field profile \(\phi(r) \propto r^{-(3w_q + 1)/2}\). This ansatz reproduces the density scaling \(\rho \propto r^{-3(w_q+1)}\) and the associated pressure components of the Kiselev fluid. For instance, in the case of quintessence with \(w_q = -2/3\), one finds \(\phi(r) \propto \sqrt{r}\) and \(V(\phi) \propto \phi^{-2}\), leading to \({T^{(\phi)}_t}^t = {T^{(\phi)}_r}^r \propto r^{-1}\), consistent with the Kiselev term \(a/r^{-1}\). This is also the desired form of quintessence potential for generating tracking solutions while enabling slow roll behavior on cosmological scales \cite{copeland2006dynamics}.

The metric function in Eq.~\eqref{eq:metric_function} captures the balance between gravitational collapse \((2M/r)\), and dark energy dominance \((a/r^{3w_q+1})\). For \(w_q \in (-1, -1/3)\), this mechanism supports acceleration driven by quintessence. Horizons are established at \(w_q = -1/3\):
\begin{equation}
    r_e = \frac{1 - \sqrt{1 - 8Ma}}{2a} ,\quad \text{and} \quad r_c = \frac{1 + \sqrt{1 - 8Ma}}{2a}.
\end{equation}
The above equation delineates a causally distinct region wherein scalar-induced anisotropy governs the dynamics \cite{ghosh2016rotating}. Stability requires that \(8Ma \leq 1\); violation of this condition results in the disappearance of horizons and the emergence of naked singularities \cite{jamil2015dynamics}. The motion of test particles in this geometry is defined by the non-zero Christoffel symbols pertinent to radial motion:
\begin{equation}
\Gamma^r_{tt} = -\frac{1}{2} f(r) f'(r), \quad \Gamma^r_{rr} = -\frac{f'(r)}{2f(r)}, \quad \Gamma^\theta_{r\theta} = \Gamma^\varphi_{r\varphi} = \frac{1}{r}, \label{eq:christoffel}
\end{equation}
and for purely radial null geodesics, the radial acceleration is governed by:
\begin{equation}
\frac{d^2 r}{d\lambda^2} - \frac{1}{2} f(r) f'(r) \left( \frac{dt}{d\lambda} \right)^2 = 0, \label{eq:null_geodesic}
\end{equation}
where \(\lambda\) is the affine parameter. Notably, the acceleration \(h\), defined as \(\sqrt{h^\mu h_\mu} = {f'(r)}/{{2f(r)}^{3/2}}\), vanishes at the extremum of \(f(r)\) (i.e., where \(f'(r)=0\)). The corresponding critical radius is given by \(r_{\text{max}} = \sqrt{2M/a}\), with \(f(r_{\text{max}}) = 1 - \sqrt{8Ma}\). This demarcates the region of scalar screening (\(r < r_{\text{max}}\)) from that of unscreened quintessence dominance (\(r > r_{\text{max}}\)), a transition central to chameleon-like behavior. Although negative pressure is required for quintessence dark energy to maintain a positive energy density (assuming \(a>0\)), the equivalence of Eqs.~\eqref{eq:tphi1}--\eqref{eq:emtensor} to the stress-energy components of a perfect fluid is not guaranteed, particularly in an orthogonal frame \cite{visser2020kiselev}. This issue will be addressed subsequently. 

Energy conservation for test particles is derived from the timelike Killing vector \(\xi^\mu = (\partial_t)^\mu\), leading to \(\mathcal{E} = f(r) ({dt}/{d\lambda})\), and the radial geodesic equation becomes \(\left({dr}/{d\lambda}\right)^2 = \mathcal{E}^2 - \epsilon f(r), \) where \(\epsilon = 1\) for massive particles and \(\epsilon = 0\) for photons. Consequently, the effective potential governing radial motion can be defined via: 
\begin{equation}
    \left(\frac{1}{2}\right) \dot{r}^2 + V_{\text{eff}}(r) = \mathcal{E},
\end{equation} where:
\begin{equation}
V_{\text{eff}}(r) = \mathcal{E} - \frac{1}{2} f(r) \left(1 - \frac{\epsilon f(r)}{\mathcal{E}^2}\right).
\label{eq:effective_potential}
\end{equation}
This relationship outlines the conditions under which particles can escape (i.e., \(\mathcal{E} > V_{\text{eff}}\)) and illustrates how anisotropic stress modifies the dynamics. In particular, the effective mass of the scalar field, as determined by its radial gradient, contributes an additional term to the effective potential. For massive particles:
\begin{equation}
    V_{\text{eff, massive}} = -\frac{M}{r} - \frac{\mathcal{K}_\phi}{r^2}, \quad \text{and for photons,} \quad V_{\text{eff, photon}} = -\frac{\mathcal{K}_\phi}{r^2},
\end{equation}where \(\mathcal{K}_\phi\) quantifies the anisotropic stress contribution (analogous to an effective charge) from the scalar field’s kinetic energy density. Explicitly, \(\mathcal{K}_\phi\) is derived from the scalar field’s gradient as:
\begin{equation}
\mathcal{K}_\phi = \frac{1}{2} \omega(\phi) \left(r \frac{d\phi}{dr}\right)^2,
\label{eq:K_phi}
\end{equation}
with \(\omega(\phi)\) governing the kinetic structure of \(\phi(r)\). For instance, in the case of quintessence with \(w_q = -2/3\), the scalar field profile \(\phi(r) \propto \sqrt{r}\) yields \(\mathcal{K}_\phi \propto r\), introducing an additional term \(\propto 1/r\) in the effective potential. This scalar contribution is distinct from the metric’s '\(a r\)' term in \(f(r)\), which governs dark energy dominance at large \(r\). 

\begin{figure}
    \centering
    \includegraphics[width=0.6\linewidth]{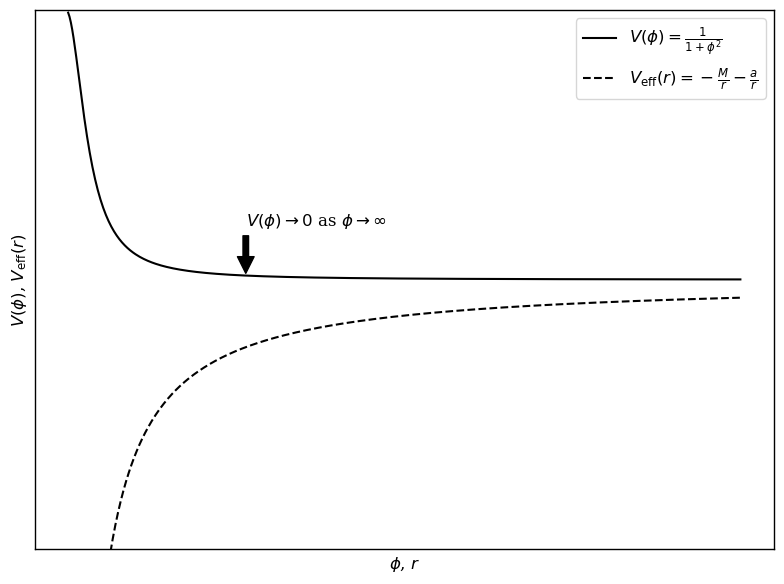}
    \caption{\small Behaviour of \(V(\phi)\) and the effective potential \(V_{\text{eff}}(r)\), asymptotically approaching \(V(\phi) \to 0\) as \(\phi \to \infty\). The scalar field’s spatial gradients modify \(V_{\text{eff}}(r)\) through anisotropic stress terms, creating a scalar-induced potential well that depends on the coupling strength \(\omega(\phi)\) and the metric’s anisotropy parameter \(w_q\). For \(\phi \gg 1\), the vanishing potential \(V(\phi)\) indicates the dominance of kinetic energy in the scalar field’s stress-energy tensor.}
    \label{fig:behaviourpotential}
\end{figure}

Physically, \(\mathcal{K}_\phi\) mimics a Reissner-Nordström-like term but is sourced by spatial gradients of scalar fields instead of electromagnetic charge. The spatial variation in \(\phi\) distinguishes the Kiselev fluid from an ideal perfect fluid. By judiciously selecting a non-minimal coupling and tuning the scalar potential (for example, via \(V(\phi) \propto \phi^n\)), one may derive a scalar field profile \(\phi(r) \propto r^{-(3w_q+1)/2}\) that reproduces the Kiselev fluid's density scaling \(\rho \propto r^{-3(w_q+1)}\) and the corresponding pressure components. A scalar field with a runaway potential \(V(\phi)\) may evolve toward vanishing energy density at large distances \cite{caldwell2005limits}, asymptotically approaching a perfect fluid regime as \(V(\phi) \to 0\) (see Figure \ref{fig:behaviourpotential}). However, residual anisotropic stresses from \(\mathcal{K}_\phi/r^2\) persist in this limit unless \(\omega(\phi)\) is specifically constrained to suppress gradient terms. Nonetheless, additional anisotropic stress terms or extra density components may be required to fully capture the characteristics of a non-minimally coupled scalar field \cite{holden2000self}, which we will discuss in subsequent sections.

\section{Scalar Field as an Anisotropic Fluid}

The dynamics of dark energy can be modeled either as a perfect fluid or using a scalar field with a runaway potential \(V(\phi)\) \cite{amendola2005constraints}. While the fluid approximation from the previous section offers a macroscopic description, it necessarily abstracts away the underlying field dynamics because the scalar field \(\phi\) and its derivative \(\partial_\mu \phi\) cannot be uniquely reconstructed from the macroscopic quantities \(\rho\) and \(p_r\) that characterize a perfect fluid \cite{babichev2005accretion}. Notwithstanding this limitation, the anisotropic stress generated by spatial gradients of \(\phi\) is a distinctive feature in this model that plays a significant role in quantifying both scalar dark energy and compact object dynamics, a phenomenon not well captured by perfect-fluid models.

In practice, when the scalar field is treated as a perfect fluid, one often assumes the absence of particle contributions, which leads to an energy-momentum tensor (EMT) with an EoS that can be difficult to interpret \cite{babichev2013black}. For \(V \neq 0\), the energy flux component of the EMT, given by \(T_0^r = -\frac{(2M)^2 \dot{\phi}_\infty^2}{r^2}\), is assumed to reflect a constant density and flow velocity over time, albeit with radial dependence. In a special case \(V = 0\), this simplifies to:
\begin{equation}
    p_\infty = \frac{\dot{\phi}_\infty^2}{2}, \quad \rho_\infty = \frac{\dot{\phi}_\infty^2}{2}.
\end{equation}
Thus, yielding identical contributions from pressure and density asymptotically.

For a runaway potential—defined by the asymptotic conditions \(\lim_{r\to\infty} V=0\), \(\lim_{r\to\infty} \frac{V_{,r}}{V}=0\), \(\lim_{r\to\infty} \frac{V_{,rr}}{V_{,r}}=0,\dots\) and by divergence near the origin, i.e., \(\lim_{r\to0} V = \infty\), \(\lim_{r\to0} \frac{V_{,r}}{V} = \infty\), \(\lim_{r\to0} \frac{V_{,rr}}{V_{,r}} = \infty,\dots\)—scalar fields in strong gravitational regimes satisfy the effective action:
\begin{equation}
    S=\int d^4 x \sqrt{-g} \left\{\frac{M_{pl}^2}{2} R - \frac{1}{2}(\partial \phi)^2 - V(\phi, r) \right\} - \int d^4x \, \mathcal{L}(\psi_{m}^{(i)}, g_{\mu \nu}^{(i)}) \label{eq:action1}
\end{equation}
where \(M_{pl} = (8 \pi G)^{-1/2}\) is the reduced Planck mass (in our geometric units, \(M_{\rm pl}\) has dimensions of [length]\(^{-1}\)), \(g\) is the determinant of the metric \(g_{\mu \nu}\), \(R\) is the Ricci scalar, and \(\psi_{m}^{(i)}\) denote the matter fields. Note that the action is employed in chameleon cosmology \cite{khoury2013chameleon} and is also the desired form for quintessence scenarios \cite{zlatev1999quintessence}.

\begin{figure}
    \centering
    \includegraphics[width=\linewidth]{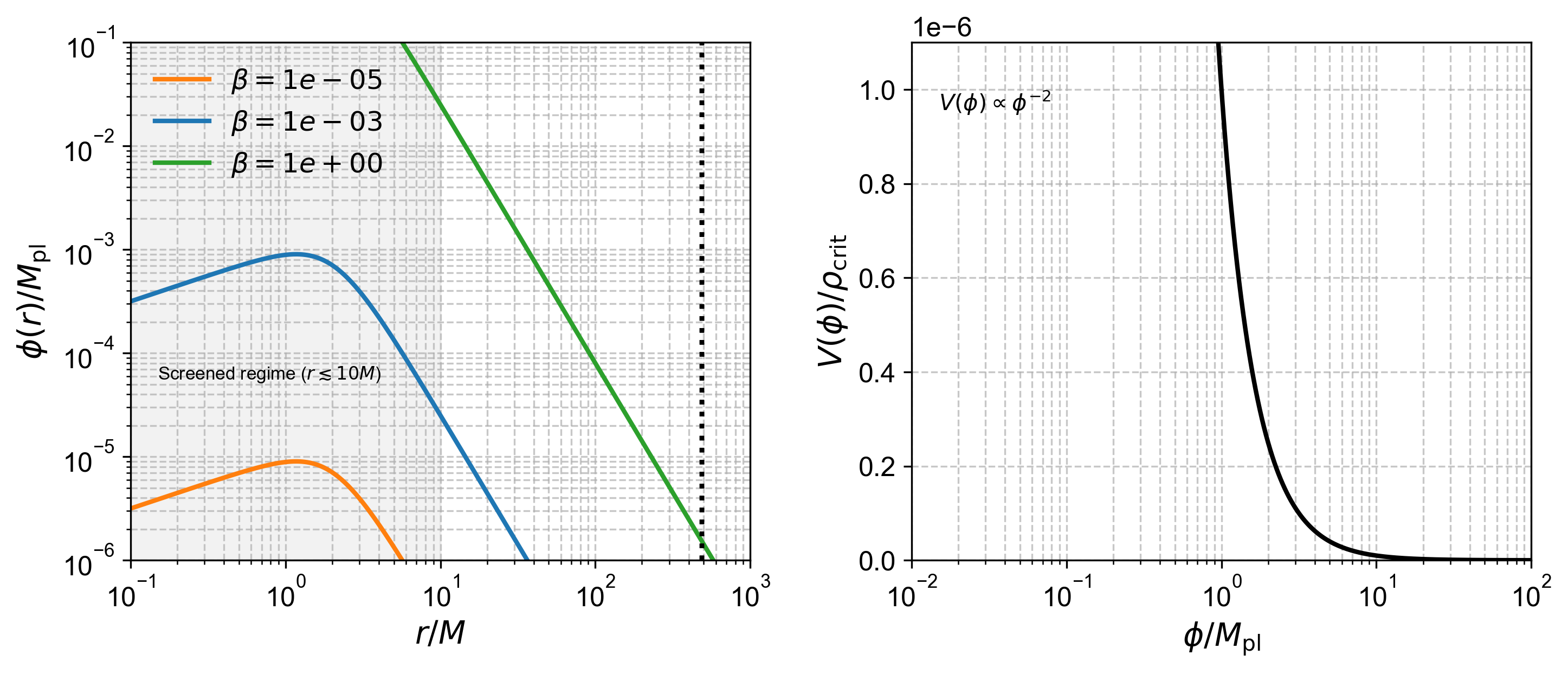}
    \caption{ \small Left panel: Log-log plot of the scalar field profile \(\phi(r)/M_{\rm pl}\) versus normalized radius \(r/M\) for three different coupling strengths.  Right panel: Semi-logarithmic plot of the runaway potential \(V(\phi)\) (scaled by a constant) versus \(\phi/M_{\rm pl}\). For small \(\beta\), the field remains suppressed over a broad radial range, while higher \(\beta\) values yield larger field amplitudes until screening takes over. The scalar field’s influence is minimal in the stellar interior; the object remains close to a GR-like solution. In contrast, as \(\beta\) increases, the field grows quickly in the outer regions before being cut off by the screening factor at large \(r \to r_{\text{max}}\). This indicates a strong coupling scenario where dark energy–like effects are locally enhanced yet remain consistent with the need for chameleon suppression in high-density regions. }
    \label{fig:scalar_profile}
\end{figure}

Another key aspect of the model is the scalar’s coupling to matter via the conformal rescaling \( g_{\mu \nu}^{(i)} = \exp\left(\frac{2 \beta_i \phi}{M_{pl}}\right) g_{\mu \nu}\). This conformal rescaling defines the Jordan-frame metric \(g_{\mu\nu}^{(i)}\) for matter species \(i\), while \(g_{\mu\nu}\) is the Einstein-frame metric. We know that the dimensionless coupling constants \(\beta_i\) introduce a density-dependent screening mechanism. Consequently, the equation of motion for \(\phi\) becomes:
\begin{equation}
    \nabla^2 \phi = V_{,\phi} - \sum_i \frac{\beta_i}{M_{pl}} \exp\left(\frac{2 \beta_i \phi}{M_{pl}}\right) g_{(i)}^{\mu \nu} T_{\mu \nu}^{(i)}.
\end{equation}
Here, \(T_{\mu \nu}^{(i)}\) is the EMT for the \(i\)th matter component, and the runaway nature of \(V(\phi)\), with \(\lim_{\phi\to\infty} V(\phi)=0\) and \(\lim_{\phi\to0} V(\phi)=\infty\) indicates that the scalar field becomes subdominant at large scales. The scalar field’s radial profile and runaway potential in Figure~\ref{fig:scalar_profile} illustrates the model's screening near the compact object and quintessence-like behavior at cosmological distances. The asymptotic behavior \(V(\phi) \propto \phi^{-2}\), where \(V(\phi)\) decreases rapidly at large \(\phi\) satisfies the quintessence boundary conditions. In a cosmological context, such an inverse-square potential can drive late-time cosmic acceleration. However, locally (near compact objects), it can induce nontrivial scalar profiles, as seen in the left panel of Figure~\ref{fig:scalar_profile}.

In the Kiselev limit (\(a \neq 0\), \(w_q = -2/3\)), the scalar field’s stress-energy mirrors the anisotropic fluid \(T^\mu_\nu = \text{diag}(\rho, \rho, -\frac{1}{2}(3w_q + 1)\rho, -\frac{1}{2}(3w_q + 1)\rho)\), where \(-\frac{1}{2}(3w_q + 1) = \frac{1}{2}\). The modified potential acquires an additional long-range correction:
\begin{equation}
      V(\phi, r) \to V(\phi) + \mathcal{K}(\phi) \left(\frac{\partial_r \phi}{r}\right)^2. \label{eq:modifiedpoten}
\end{equation} For \(\phi(r) \propto r^{\alpha}\), this term scales as \(r^{-2}\) with the scalar fluid’s stress-energy landscape similar to those encountered in effective field theory approaches \cite{hill1988models}. Substituting Eq.~\eqref{eq:modifiedpoten} into Eq.~\eqref{eq:action1}, we yields the effective action:
\begin{dmath}
    S = \int d^4x \sqrt{-g} \left[\frac{M_{\text{pl}}^2}{2} R - \frac{1}{2}(\partial \phi)^2 - \left(V(\phi) + \mathcal{K}(\phi) \left(\frac{\partial_r \phi}{r}\right)^2\right)\right] - \int d^4x \, \mathcal{L}(\psi_m^{(i)}, g_{\mu\nu}^{(i)}), \label{eq:action2}
\end{dmath}
which introduces an additional constraint from a long-range force that influences the motion of test particles and photons in regions of strong gravitational gradients. The scalar field gradient and its coupling to the matter density can be defined using the equation:
\begin{equation}
    \nabla^2 \phi = V_{,\phi} + \frac{\mathcal{K}_{\phi,\phi}}{r^2} - \sum_i \frac{\beta_i}{M_{pl}} T^{(i)},
\end{equation} where \( \mathcal{K}_{\phi,\phi} = {\partial^2 \mathcal{K}(\phi)}/{\partial \phi^2}\) arises from the modulation of the scalar field’s kinetic energy by spacetime curvature (scaling as \(M/r^3\)) and quantifies the deviation from anisotropic Kiselev spacetime. If we assume a spherical symmetry where the spatial gradient term \((\partial_r \phi/r)^2\) reduces the full covariant kinetic term \(\partial_\mu \phi \partial^\mu \phi\) to its radial component, we obtain the standard quintessence parametrization of the Kiselev metric. For instance, if \(\mathcal{K}(\phi) = \phi^{4/3}\) and \(\phi(r) = r^{3/2}\), then \(\mathcal{K}_\phi/r^2 =\frac{9}{4} r\), matching the Kiselev term \(a/r^{3w_q + 1}\) for \(w_q = -2/3\).

For non-relativistic matter, with \(T_{\mu\nu}^{(i)} = \rho_i u_\mu u_\nu\), taking the trace and expressing it in terms of the conserved density \(\rho_i\) in the Einstein frame yields:
\begin{equation}
    \nabla^2 \phi = V_{,\phi} +\frac{\partial^2 \mathcal{K}(\phi)}{\partial \phi^2} \frac{(\partial_r \phi)^2}{r^2}  + \sum_i \frac{\beta_i}{M_{pl}} \rho_i \exp\left(\frac{3\beta_i \phi}{M_{pl}}\right), \label{eq:nabsquare}
\end{equation}
which shows that the divergence of \(V\) as \(r\to0\) stabilizes the scalar field near compact objects when \(\rho_i\) is independent of \(\phi\). This stabilization suppresses deviations from GR in the strong-field regime, leading to an effective potential of the form:
\begin{equation}
    V_{\text{eff}}(\phi) = V(\phi, r) + \sum_i \frac{\beta_i}{M_{pl}} \rho_i \exp\left(\frac{3\beta_i \phi}{M_{pl}}\right). \label{eq:effectpoten}
\end{equation}
In regions far from the black hole, where \(\dot{\phi}_\infty^2\) remains constant, one finds \(\mathcal{K}_{\phi,\phi} \to 0\), so that Eq.~\eqref{eq:nabsquare} reduces to:
\begin{equation}
    \nabla^2 \phi = V_{,\phi} + \sum_i \frac{\beta_i}{M_{pl}} \rho_i \exp\left(\frac{3\beta_i \phi}{M_{pl}}\right), \label{eq:general_eq_simplified}
\end{equation}
and a monotonic potential emerges akin to the low-energy effective actions found in string theory due to non-perturbative effects \cite{anderson1998dark, huey2000cosmological}. It is also important to note that when a compact object is surrounded by a light scalar field, a nearly constant local matter density \(\rho_m\) in the object’s core stabilizes the field and leads to small gradients. The density-dependent mass of scalar field  \(m_\phi(\rho_m)\) is defined as: \begin{equation}
m_\phi^2 = V_{,\phi\phi} + \mathcal{K}_{,\phi\phi} \left(\frac{\partial_r \phi}{r}\right)^2 + \sum_i \frac{3\beta_i^2}{M_{pl}^2} \rho_i \exp\left(\frac{3\beta_i \phi}{M_{pl}}\right),
\end{equation} 
where high \(\rho_m\) screens the scalar force, while low \(\rho_m\) allows the runaway behavior. In high-density regions, \(\mathcal{K}_\phi \propto (\partial_r \phi)^2\) dominates, enhancing \(m_\phi\) (\(\rho_m\gg V,_{\phi}\)), thereby suppressing fifth forces via the chameleon mechanism. Conversely, in low-density regions (\(\rho_m \ll V_{,\phi}\)), \(m_\phi\) decreases, allowing for long-range interactions that are consistent with quintessence-driven cosmic acceleration. Moreover, the coupling constants \(\beta_i\) are constrained by solar system tests and neutron star observations \cite{Sagunski2017Neutron} to limit deviations from GR, with fifth-force experiments requiring \(\beta_i \lesssim 10^{-3}\) for \(M_{pl} \sim 10^{18}\) GeV. This bound arises from Cassini tracking in the solar system \cite{Barreira2015K-mouflage}, where the scalar-mediated fifth force must satisfy \(\Delta G/G_N \lesssim 10^{-5}\). 

\section{Non-Minimally Coupled Scalar Field in the Vicinity of Exotic Compact Objects}

The energy-momentum tensor (EMT) of a non-minimally coupled scalar field generically violates the weak energy condition (WEC) in anisotropic regimes \cite{bekenstein1974exact, ford2001classical}, thereby enabling exotic compact objects such as boson stars and gravastars to sustain macroscopic regions of negative energy density. Although WEC violation is typically viewed as pathological within general relativity, scalar-tensor theories accommodate such configurations via spacetime-dependent screening mechanisms—for example, the chameleon effect, in which the effective mass \(m_{\phi}\) increases in high-density regions \cite{fewster2006averaged}. In this context, horizonless objects stabilized by scalar hair \cite{liebling2023dynamical, jetzer1992boson} are able to evade no-hair theorems through anisotropic stress-energy contributions that are naturally encoded in the Kiselev metric. In particular, the metric function, which behaves as \(f(r) \sim r^{-n}\) (for \(n > 0\)) near compact objects, transitions smoothly from a regime of high-density screening (\(r \ll r_{\text{crit}}\)) to one dominated by quintessence-like dynamics (\(r \gg r_{\text{crit}}\)), where the critical radius \(r_{\text{crit}}\) indicates the scalar field’s ability to screen anisotropies, with larger \(\phi_0\) or stronger couplings \(\beta_i\).

It is instructive to consider how the scalar field’s spatial gradients modify the metric function. This modification is encapsulated in the scalar-metric coupling term:
\begin{equation}
\frac{\partial f(r)}{\partial \phi} = -\frac{3\beta_i}{M_{\text{pl}}} \frac{a}{r^{3w_q + 1}} \exp\!\left(\frac{3\beta_i \phi}{M_{\text{pl}}}\right), \label{eq:derivativeoffr}
\end{equation}
where the normalization factor \(a \) is fixed by the scalar field’s amplitude \(\phi_0\) and coupling \(\beta_i\) (see Appendix A). Notably, in the limit \(\beta_i \to 0\), the metric decouples from the scalar field, reducing to the trivial case \(f(r) \to 1\) where the kinetic energy of test particles readily overcomes any potential barrier.

\begin{figure}
    \centering
    \includegraphics[width=0.8\linewidth]{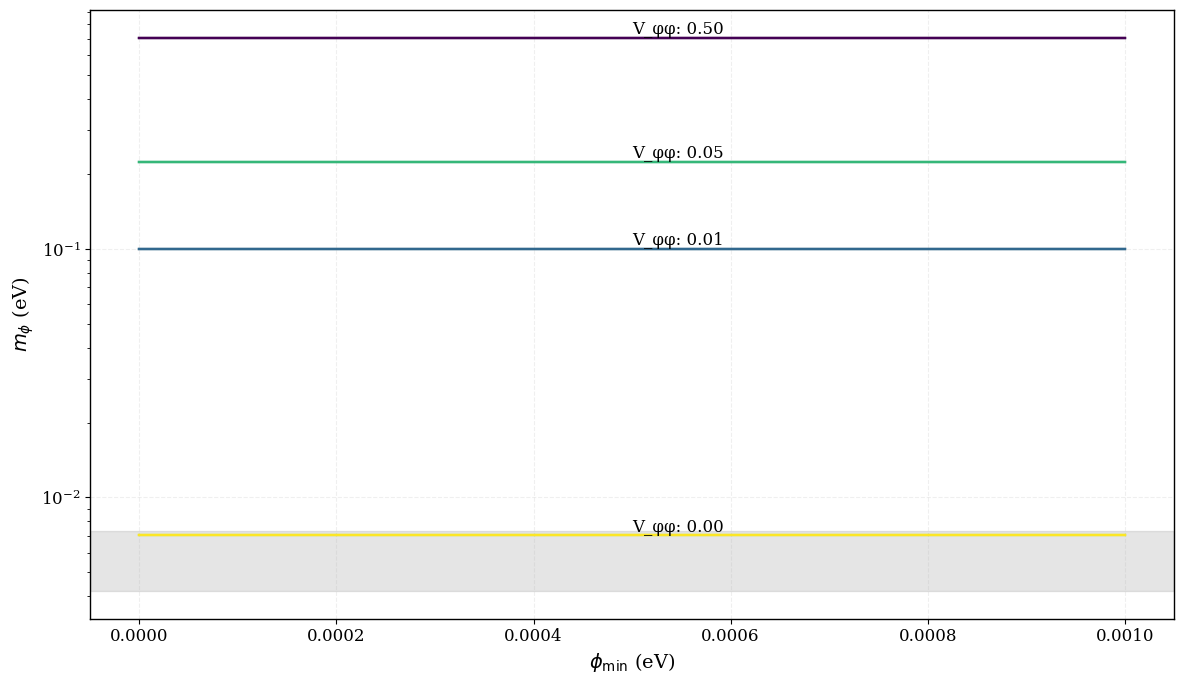}
    \caption{\small Bare scalar mass \(( m_\phi )\) as a function of the scalar field minimum \( (\phi_{\text{min}} )\). The specific range of parameter choices are: \(w_q = -2/3\),  \( a \) = \( [ 0.01 - 0.005] \), and \( \beta_i \) = \( [10^{-1} - 10^{-4}] \). The quantum stability bounds \(0.0042 < m_\phi < 0.0073\,{\rm eV}\) \cite{Upadhye2012Quantum} were marked in grey with loop corrections that maintain a stable chameleon field, while those lying outside may introduce quantum instabilities or require fine-tuning. Stability requires \(m_\phi^2 > 0\), dominated by \(V_{,\phi\phi}\) (steep potentials) and \((m_\phi^2 < 0)\) may trigger gravitational collapse into black holes or scalar halo formation, depending on the timescale of instability \cite{liebling2023dynamical}. }
    \label{fig:mvsphi}
\end{figure}

At equilibrium, the scalar field \(\phi\) attains a local minimum \(\phi_{\text{min}}\) determined by the balance of competing forces. This condition is expressed by the vanishing of the scalar field’s equation of motion:
\begin{equation}
0 = V_{,\phi} - \frac{1}{2r^2} \frac{3\beta_i}{M_{\text{pl}}} \frac{a\,\exp\!\left(\frac{3\beta_i \phi_{\text{min}}}{M_{\text{pl}}}\right)}{r^{3w_q + 1}} + \sum_i \frac{\beta_i}{M_{\text{pl}}} \rho_i\,\exp\!\left(\frac{3\beta_i \phi_{\text{min}}}{M_{\text{pl}}}\right).
\end{equation}
Here, \(V_{,\phi}\) denotes the derivative of the background scalar field potential, and the summation term accounts for the contributions from various density components \(\rho_i\) with corresponding couplings \(\beta_i\). To further elucidate the dynamics, one linearizes the field around equilibrium by writing \(\phi = \phi_{\text{min}} + \delta \phi\), where \(\delta \phi\) represents small perturbations induced, for instance, by a localized point mass. This linearization yields:
\begin{equation}
\nabla^2 \delta \phi = \left. V_{,\phi\phi} \right|_{\phi_{\text{min}}} \delta \phi + \frac{1}{2r^2} \left. \frac{\partial^2 f(r)}{\partial \phi^2} \right|_{\phi_{\text{min}}} \delta \phi + \sum_i \frac{\beta_i}{M_{\text{pl}}} \rho_i \frac{3\beta_i}{M_{\text{pl}}} \exp\!\left(\frac{3\beta_i \phi_{\text{min}}}{M_{\text{pl}}}\right) \delta \phi, \label{eq:eomscalar}
\end{equation}
from which the scalar mass \(m_\phi\) is naturally identified via
\begin{equation}
m_\phi^2 = \left. V_{,\phi\phi} \right|_{\phi_{\text{min}}} + \frac{1}{2r^2} \left. \frac{\partial^2 f(r)}{\partial \phi^2} \right|_{\phi_{\text{min}}} + \sum_i \frac{\beta_i}{M_{\text{pl}}} \rho_i\,\frac{3\beta_i}{M_{\text{pl}}} \exp\!\left(\frac{3\beta_i \phi_{\text{min}}}{M_{\text{pl}}}\right). \label{eq:scalarmass}
\end{equation}
The metric-dependent term \({\partial^2 f(r)}/{\partial \phi^2}\) encodes the scalar’s backreaction on the geometry through \(a(\phi_0, \beta_i)\) as derived in Appendix A. Expanding further using the expression in Eq.~\eqref{eq:derivativeoffr}, one obtains:
\begin{dmath}
m_\phi^2 = \left. V_{,\phi\phi} \right|_{\phi_{\text{min}}} - \frac{(3\beta_i)^2}{2r^2\,r^{3w_q + 1}\,M_{\text{pl}}^2} a\,\exp\!\left(\frac{3\beta_i \phi_{\text{min}}}{M_{\text{pl}}}\right) + \sum_i \frac{\beta_i}{M_{\text{pl}}} \rho_i\,\left(\frac{3\beta_i}{M_{\text{pl}}}\right) \exp\!\left(\frac{3\beta_i \phi_{\text{min}}}{M_{\text{pl}}}\right). \label{eq:expansion}
\end{dmath}
This expression, while mathematically intricate, has a clear physical interpretation: in regimes where the intrinsic potential \(V(\phi)\) is steep, the term \(V_{,\phi\phi}\) dominates, determining \(m_\phi\); however, when the potential is flatter, the metric-dependent coupling terms, which incorporate the effects of matter density and anisotropic corrections, can significantly modify \(m_\phi\) and even lead to negative-energy fluctuations that trigger further perturbations in the scalar field. In the asymptotic limit \(w_q \to -1\), Eq.~\eqref{eq:expansion} simplifies to:
\begin{equation}
m_\phi^2 = \left. V_{,\phi\phi} \right|_{\phi_{\text{min}}} - \frac{(3\beta_i)^2}{2M_{\text{pl}}^2} a\,\exp\!\left(\frac{3\beta_i \phi_{\text{min}}}{M_{\text{pl}}}\right) + \sum_i \frac{\beta_i}{M_{\text{pl}}} \rho_i\,\left(\frac{3\beta_i}{M_{\text{pl}}}\right) \exp\!\left(\frac{3\beta_i \phi_{\text{min}}}{M_{\text{pl}}}\right).
\end{equation}
In this scenario, the effective scalar mass is predominantly governed by the curvature of the intrinsic potential, while the additional contributions from matter and dark energy density, modulated by the metric-dependent couplings, serve as corrections. Figure~\ref{fig:mvsphi} illustrates this dependence for various parameter choices.

It is important to note that the scalar mass \(m_\phi\) shows significant sensitivity to matter interactions, making its direct detection difficult.  For example, compact objects such as neutron stars (PSR J0740+6620 \cite{riley2021nicer, Collaboration2018Properties}) require \(m_\phi > 10^{-22}\) eV in order to produce observable deviations in gravitational wave signals, consistent with constraints from LIGO-Virgo measurements \cite{Berti2012Light, Sagunski2017Neutron}. Moreover, scalar hair modifies the tidal Love numbers \(k_2\) of neutron stars, which leaves distinct signatures in the inspiral waveform for scalar couplings \(\beta_i \sim 10^{-3}\). For binary neutron star mergers (GW170817 \cite{Abbott2017GW170817:}), LIGO-Virgo constrains the dimensionless tidal deformability \(\tilde{\Lambda}_{1.4M_\odot} \leq 580\), which scalar hair reduces by \(\sim 10\)–\(30\%\) \cite{sennett2017distinguishing}. This is expected because the scalar-mediated forces effectively stiffen the EoS, thereby suppressing scalar effects in the dense core while restoring GR-like behavior. In the unscreened outer layers, the scalar field becomes active, adds to the total gravitational source, and acts inward, increasing gravitational binding energy.

\section{Screened-Unscreened Transition: Quasi-static Profiles of Compact Objects}

In spherical symmetry, the scalar field must satisfy \(d\phi/dr \to 0\) at the inner boundary \(r_e\) and asymptote to its cosmological value \(\phi \to \phi_\infty\) at the outer boundary \(r_c\). Accordingly, the scalar force for a maximally symmetric Kiselev metric, as expressed in Eq.~(\ref{eq:nabsquare}), takes the radial form:
\begin{equation}
\frac{d^2 \phi}{dr^2} + \frac{2}{r} \frac{d\phi}{dr} = V_{,\phi} + \frac{\mathcal{K}_{\phi, \phi}}{r^2} + \sum_i \frac{\beta_i}{M_{\text{pl}}} \rho_i(r) \exp\!\left(\frac{3\beta_i \phi}{M_{\text{pl}}}\right). \label{eq:sphericalsymmetry}
\end{equation}
The above equation must be solved in the domain \(r_e < r < r_c\), where its solution is constrained by the boundary conditions matching the Kiselev metric. At \(r \to r_c\) (the cosmological horizon), we require \(\phi \to \phi_\infty\), \(\rho \to \rho_\infty\), and \(d\phi/dr \to 0\); while at \(r \to r_e\) (near the event horizon), the solution must align with the anisotropic stress components of the EMT, effectively suppressing the scalar force in that region \cite{betz2022searching}. Stability analysis of Eq.~(\ref{eq:sphericalsymmetry}) is provided in Appendix B.

In order to compute the scalar force \(F_\phi\) at a given radial distance \(r\), we consider the integrated form of the radial equation by introducing a dummy integration variable \(r'\). Specifically, the scalar field solution is approximated as:
\begin{equation}
\phi(r) \approx \phi_{\infty} - \int_r^\infty \left[ V_{,\phi} + \frac{\mathcal{K}_{\phi, \phi}}{r^{'2}} + \sum_i \frac{\beta_i}{M_{\text{pl}}} \rho_i(r') \exp\!\left(\frac{3\beta_i \phi(r')}{M_{\text{pl}}}\right) \right] dr'. \label{eq:dummyfun}
\end{equation}
To render Eq.~(\ref{eq:dummyfun}) tractable, we first linearize the scalar potential \(V(\phi)\) around small field values, adopting a quadratic form \(V(\phi)=\frac{1}{2} m^2 \phi^2\) where \(V_{,\phi}=m^2 \phi\). Next, we assume the relative density variation \(\Delta \rho/\rho_{\infty} \sim \mathcal{O}(1)\), with the density profile specified as compact star density gradients \(\rho_i(r)=\rho_{\infty}\left(1-\frac{r_c-r}{r_c-r_e}\right)\), such that \(\rho_i(r_e)\approx 2\rho_{\infty}\) mimics crust-to-core (two-piece) transitions. Furthermore, for small field values, the exponential term undergoes a perturbative treatment \((\exp\!\left(\frac{3\beta_i \phi}{M_{\text{pl}}}\right) \approx 1 + \frac{3\beta_i \phi}{M_{\text{pl}}} + \dots)\) and holds for \(\phi \ll M_{\text{pl}}/\beta_i\), where higher-order terms (\(\sim \phi^2\)) remain negligible. Consequently, Eq.~(\ref{eq:dummyfun}) becomes:
\begin{equation}
\phi(r) \approx \phi_{\infty} - \int_r^\infty \left[ \left(m_{\phi}^2 + \sum_i \frac{3\beta_i^2 \rho_i(r')}{M_{\text{pl}}^2}\right) \phi + \frac{\mathcal{K}_{\phi, \phi}}{r^{'2}} + \sum_i \frac{\beta_i \rho_i(r')}{M_{\text{pl}}} \right] dr', \label{eq:fullequ}
\end{equation}
which can be solved by decomposing the solution into homogeneous and particular parts. 

The homogeneous part satisfies the differential equation:
\begin{equation}
\phi_h'' + \frac{2}{r} \phi_h' - m_{\text{eff}}^2 \phi_h = 0,
\end{equation}
where the effective mass is defined by:
\begin{equation}
m_{\text{eff}}^2 = m_{\phi}^2 + \sum_i \frac{3\beta_i^2 \rho_{\infty}}{M_{\text{pl}}^2}.
\end{equation}
For quasi-static density profiles (\(\Delta \rho/\rho_\infty \sim \mathcal{O}(1)\)), \(m_{\text{eff}}^2\) is approximated using \(\rho_\infty\), whereas for strongly varying \(\rho_i(r)\), we replace \(\rho_\infty \to \rho_i(r)\). The general solution to this homogeneous equation is a Yukawa-type decay:
\begin{equation}
\phi_h(r) = C_1 \frac{\exp\!\left(-{m_{\text{eff}}^2}\, r\right)}{r},
\end{equation}
with $C_1$ determined by matching to compact object boundary conditions. Note that the homogeneous solution diverges as \(r \to 0\), violating finite-energy requirements for compact objects. Thus, \(C_1 = 0\). This approximation holds well in regions where matter density varies slowly. However, near the crust–core boundary of neutron stars or in other environments where density gradients are sharp, the assumption of a constant \(m_{\rm eff}\) may break down, and hence the Yukawa–like solution could deviate from the true behavior of the field.

The particular solution, incorporating anisotropic stresses and matter couplings, is obtained via the Green's function formalism:
\begin{equation}
    \phi_p(r) = \int_0^\infty G(r, r') \left( \frac{\mathcal{K}_{\phi,\phi}}{r'{^2}} - \sum_i \frac{\beta_i \rho_i(r')}{M_{\text{pl}}} \right) dr', 
    \label{eq:particular}
\end{equation}
where $G(r, r') = \frac{e^{-m_{\text{eff}} |r - r'|}}{4\pi |r - r'|}$ is the Yukawa Green's function which assumes \(m_{\text{eff}}^2\) varies slowly over screening scales (\(r_{\text{crit}} \sim m_{\text{eff}}^{-1}\)), consistent with quasi-static density profiles in Section 3. For $r \gg m_{\text{eff}}^{-1}$, this reduces to:
\begin{equation}
    \phi_p(r) \approx \frac{\mathcal{K}_{\phi,\phi}}{r} - \sum_i \frac{\beta_i}{M_{\text{pl}}} \int_r^\infty \frac{\rho_i(r')}{r'} e^{-m_{\text{eff}}(r' - r)} dr'. 
    \label{eq:particular_approx}
\end{equation}

Imposing asymptotic flatness ($\phi \to \phi_\infty$ as $r \to \infty$), the full solution becomes:
\begin{equation}
    \phi(r) = \phi_\infty + \frac{\mathcal{K}_{\phi,\phi}}{r} - \sum_i \frac{\beta_i}{M_{\text{pl}}} \int_r^\infty \frac{\rho_i(r')}{r'} e^{-m_{\text{eff}}(r' - r)} dr', 
    \label{eq:full_solution}
\end{equation}
where the homogeneous solution vanishes ($C_1 = 0$) to ensure finite energy and the anisotropic stress term \(\mathcal{K}_{\phi,\phi}/r^2\) corresponds to the tangential pressure deficit \(T^\theta_\theta - T^r_r\). Differentiating Eq. \eqref{eq:full_solution} with respect to \(\phi\)  yields the radial scalar force:
\begin{equation}
    F_\phi = -\frac{d\phi}{dr} = \frac{\mathcal{K}_{\phi,\phi}}{r^2} - \sum_i \frac{\beta_i \rho_i(r)}{M_{\text{pl}}} e^{-m_{\text{eff}} r} + \mathcal{O}\left(\frac{e^{-m_{\text{eff}} r}}{r^2}\right), 
    \label{eq:scalarfor}
\end{equation}
exhibiting two distinct regimes: (i) $r \ll m_{\text{eff}}^{-1}$, where the anisotropic stress term $\mathcal{K}_{\phi,\phi}/r^2$ dominates, and (ii) $r \gg m_{\text{eff}}^{-1}$, governed by Yukawa-suppressed matter couplings. Note that, in section 2,  \(\mathcal{K}_{\phi,\phi} \equiv {\partial^2 \mathcal{K}(\phi)}/{\partial \phi^2}\) can be negative if \(\mathcal{K}(\phi)\) is concave. However, the sign of \(\mathcal{K}_{\phi,\phi}\) in Eq. \eqref{eq:scalarfor} indicates the curvature of \(\mathcal{K}(\phi)\) is convex \(\mathcal{K}_{\phi,\phi} < 0\) that would enhance screening.

At the intermediate radius \(r_{\text{crit}}=\sqrt{2M/a}\), the maximum scalar force \(F_{\phi,\text{max}}\) arises where the two contributions balance. At small distances (\(r_{\text{crit}} \ll \sqrt{2M}\)) the force primarily originates from the anisotropic stress term, with \(F_{\phi,\text{max}} \approx \frac{\mathcal{K}_{\phi, \phi}\, a^2}{2M}\), thereby breaking the balance between radial and tangential pressures. By ignoring the higher-order term (since it's small near \(r_{\text{crit}}\)), Eq. \eqref{eq:scalarfor} becomes:
\begin{equation}
    \frac{\mathcal{K}_{\phi,\phi}}{r^2} = \sum_i \frac{\beta_i \rho_i(r)}{M_{\text{pl}}} e^{-m_{\text{eff}} r}.\label{eq:balanceeqn}
\end{equation}the critical radius \(r_{\text{crit}} \equiv \sqrt{2M/a}\) corresponds to the screened-unscreened transition scale defined in Section 1, balancing scalar anisotropy and matter couplings. However, phenomenological density profiles are essential to accurately determine \(r_{\text{crit}}\) and to test against LIGO-Virgo tidal deformability bounds (\(\tilde{\Lambda}_{1.4M_\odot} \leq 800\)) as in Section 3. In the following section, we demonstrate the analysis of neutron stars as compact objects through distinct couplings (\(\beta_i = \beta\)) and density profiles ($\rho_i = \rho$).
\section{Neutron Star Structure in Scalar-Tensor Gravity with Anisotropic Stresses}
\label{sec:neutron_stars}

The hydrostatic equilibrium of neutron stars in the presence of a non-minimally coupled scalar field is governed by the generalized Tolman-Oppenheimer-Volkoff (TOV) equations \cite{campitelli2024neutron}:
\begin{equation}
    \frac{dp_r(r)}{dr} = -\frac{\left(p_r+\rho+\rho_q\right)\left(m+4\pi r^3\left(p_r-\rho_q\right)\right)}{r\left(r-2m\right)} + \frac{2\Delta p}{r}, \label{eq:mod_tov}
\end{equation}
\begin{equation}
    \frac{dm}{dr} = 4\pi r^2 \left(\rho+\rho_q\right), \label{eq:mass_eq}
\end{equation}
where the anisotropic stress is quantified by \(\Delta p = p_r - p_t = \left(\partial_r\phi\right)^2 \geq 0,\) and the scalar field contributes an energy density \(\rho_q = \frac{1}{2}\left(\partial_r\phi\right)^2 + V(\phi),\) with a quartic order scalar potential\cite{odintsov2022neutron}:
\begin{equation}
    V(\phi)=V_0\,\phi^{-2}.
\end{equation}
In the absence of scalar fields (\(\rho_q=0\) and \(p_r=p_t\)) the above equations reduce to the standard TOV system. However, the nonzero spatial gradients \(\partial_r\phi\) and the potential \(V(\phi)\) introduce nonlinear modifications to both the pressure balance and the mass profile. Note that we use geometric units \((G=c=1)\) throughout the section.

Neutron stars modeled by these equations exhibit extreme core densities (\(\rho_c \sim 10^{15}\,\mathrm{g/cm^3}\)) and steep density gradients \cite{glendenning2000lower}. For the sake of simplicity and to illustrate the effects of the scalar field, we adopt a polytropic equation of state of the form: 
\begin{equation}
    P=K_{}\,\rho^\gamma, \label{eq:eospoly}
\end{equation}which may be inverted to yield the density as a function of pressure, \(\rho=(P/K_{})^{1/\gamma}\). In our numerical integrations, the polytropic index \(\gamma\) and constant \(K\) are chosen to mimic the qualitative behavior of nuclear matter. To effectively absorb the outer-layer (crustal) behavior, we tune \(K\) and \(\gamma\) such that the resulting M–R curve respects the tidal deformability and radius constraints from NICER and LIGO–Virgo observations without requiring an explicit crust model. While we adopt the polytropic approximation for modeling purposes, distinct EoSs such as the WFF1 \cite{wiringa1988equation}, which is a variational method EoS, SLy\cite{douchin2001unified}, which is a potential method EoS, and the APR EOS \cite{akmal1998equation} provide more microphysical realism. This should be accounted for in future work while computing observational bounds on the model to real-time compact object data.

The scalar field is assumed to follow an ansatz used in \cite{crawford2009neutron}. For instance, one may assume \(\phi(r)=\phi_0\,r^{1/2},\) where \( \phi_0 \) is fixed by boundary conditions at the neutron star surface so that its radial derivative is: 
\begin{equation}
    \partial_r\phi=\frac{\phi_0}{2}\,r^{-1/2} \quad \Longrightarrow \quad \left(\partial_r\phi\right)^2=\frac{\phi_0^2}{4\,r}.
\end{equation}
With this prescription, the scalar energy density becomes:
\begin{equation}
    \rho_q(r)=\frac{1}{2}\left(\partial_r\phi\right)^2+V(\phi)
    =\left(\frac{\phi_0^2}{8\,}+\frac{V_0}{\phi_0^2 }\right)r^{-1},
\end{equation} and the anisotropic pressure is simply \(\Delta p= \rho_q/2={\phi_0^2}/{4\,r}\) matches the Kiselev fluid’s tangential pressure deficit (Section 1) for \(w_q = -2/3\). The coupling to matter fixed by the normalization factor is defined by: 
\begin{equation}
    a=\frac{8\pi \phi_0^2}{3}\,\exp\left(\frac{3\beta\phi_0}{M_{\text{pl}}}\right),
\end{equation}
which determines the transition between screened (\(r<r_{\text{crit}}\)) and unscreened (\(r>r_{\text{crit}}\)) regimes (See Appendix A). By substituting \(\rho_q\) into a scalar–matter equilibrium condition (Eq. \eqref{eq:balanceeqn}), we get:
\begin{equation}
    \frac{a\,\rho_c}{r^2}=\frac{\beta\,\rho_c\,r_{\text{core}}}{M_{\text{pl}}\,r}\,e^{-m_{\text{eff}}r}, \quad m_{\text{eff}}^2=m_\phi^2+\frac{3\beta^2\rho}{M_{\text{pl}}^2},
\end{equation} where \(\rho_c\) is the central density of the star satisfying the boundary condition \(p_r(0)=p_c \), \(p_r(R)=0\), \(m(R) = M_{NS}\) and \( \phi(R) = \phi_0 R^{-1/2} \). This yields the critical screening radius (in geometric units) through a perturbative expansion:
\begin{equation}
    r_{\text{crit}}^{(0)}=\sqrt{\frac{a\,M_{\text{pl}}}{\beta\,r_{\text{core}}}}, \quad r_{\text{crit}}^{(n+1)}=r_{\text{crit}}^{(n)}\left[1+\frac{m_{\text{eff}}\,r_{\text{crit}}^{(n)}}{2}\right].\label{eq:criticalradius}
\end{equation}We computed initial estimates for two sets of parameters in geometric units: for \(\phi_0=10^{-3},\, \beta = 10^{-3}, \, m_{\text{eff}}=10^{-6} \,\text{eV} \) (heavy scalar field) and \(\phi_0=10^{-3}, \, \beta = 10^{-3}, \, m_{\text{eff}}=10^{-24} \,\text{eV} \) (ultra-light scalar field) (see Appendix C). In both cases, the initial value is around \(r_{\text{crit}}^{(0)}\approx 0.02894  \,\text{km}\). The iterative procedure rapidly converges to the physical value of the critical radius as shown in Figure \ref{fig:iteration} such that the scalar contributions become significant only in a more confined outer region of the star for lower \(\phi_0\). We intend to discuss this in detail in section 8.

Although realistic neutron star models employ piecewise EoS to capture the crust–core transition \cite{read2009constraints, read2009measuring}, in the present study, we adopt a single relativistic polytropic EoS defined in Eq. \eqref{eq:eospoly}. This approximation captures the qualitative behavior of nuclear matter at high densities, which are most relevant for probing scalar–tensor modifications that adopt single-polytropic models for neutron star interiors \cite{damour1996tensor, silva2016low}. While the outer crust is not explicitly modeled, its physical influence is effectively absorbed into the choice of polytropic parameters. In particular, we calibrate the constants $K$ and $\gamma$ such that the resulting neutron star configurations with $M = 1.4\,M_\odot$ yield radii consistent with current observational bounds ($R_{1.4} \lesssim 11.9\,\text{km}$) as reported by Capano et al. \cite{capano2020stringent}. The modified TOV equations (Eqs.~\eqref{eq:mod_tov} and \eqref{eq:mass_eq}) are integrated numerically using a Runge–Kutta method, starting from a small inner radius $r_0$ to avoid coordinate singularities. At $r = r_0$, the central pressure $p_c$ is related to the central density $\rho_c$ via the polytropic relation $p_c = K_{\text{ST}} \rho_c^\gamma$, and the initial enclosed mass is set to zero, $m(r_0) = 0$. The integration proceeds outward until the pressure drops to zero, thereby defining the stellar radius $R$.

At the surface, the numerical interior solution for the metric functions and scalar field, along with their radial derivatives, are matched continuously to the analytic exterior solution obtained in isotropic coordinates by \cite{crawford2009neutron}. The exterior solution is transformed to Schwarzschild coordinates for matching, preserving continuity of \(g_{\mu, \nu}\) and \(\phi\). The analytic exterior solution is given by:
\begin{equation}
    e^{2\alpha(r)}=e^{2\alpha_\infty}\left(\frac{r-r_0}{r+r_0}\right)^{2A}, \quad e^{2\theta(r)}=e^{2\theta_\infty}\left(1-\frac{r_0^2}{r^2}\right)^2\left(\frac{r-r_0}{r+r_0}\right)^{-2A},
\end{equation}
\begin{equation}
    \phi(r)=\sqrt{\frac{2}{\kappa}}\sqrt{1-A^2}\,\ln\left(\frac{r-r_0}{r+r_0}\right)+\phi_\infty,
\end{equation}
where the integration constants \(r_0\), \(A\), \(\alpha_\infty\), \(\theta_\infty\), and \(\phi_\infty\) are determined by demanding continuity of \(\alpha\), \(\theta\), \(\phi\) and their derivatives at \(r=R\). By setting \(\alpha_\infty=\theta_\infty=0\) (which amounts to a rescaling of the time and radial coordinates), the matching conditions reduce to equating the interior functions \(\alpha_{\rm int}(R)\), \(\theta_{\rm int}(R)\), and \(\phi_{\rm int}(R)\) and their derivatives with the corresponding exterior expressions. In particular, the continuity of the derivative of \(\alpha(r)\) at \(r=R\) provides an algebraic relation between the constant \(A\) and \(r_0\), while the derivative of \(\phi(r)\) yields the value of \(\sqrt{1-A^2}\). Finally, the continuity of the scalar field itself fixes \(\phi_\infty\).

\begin{figure}
    \centering
    \includegraphics[width=\linewidth]{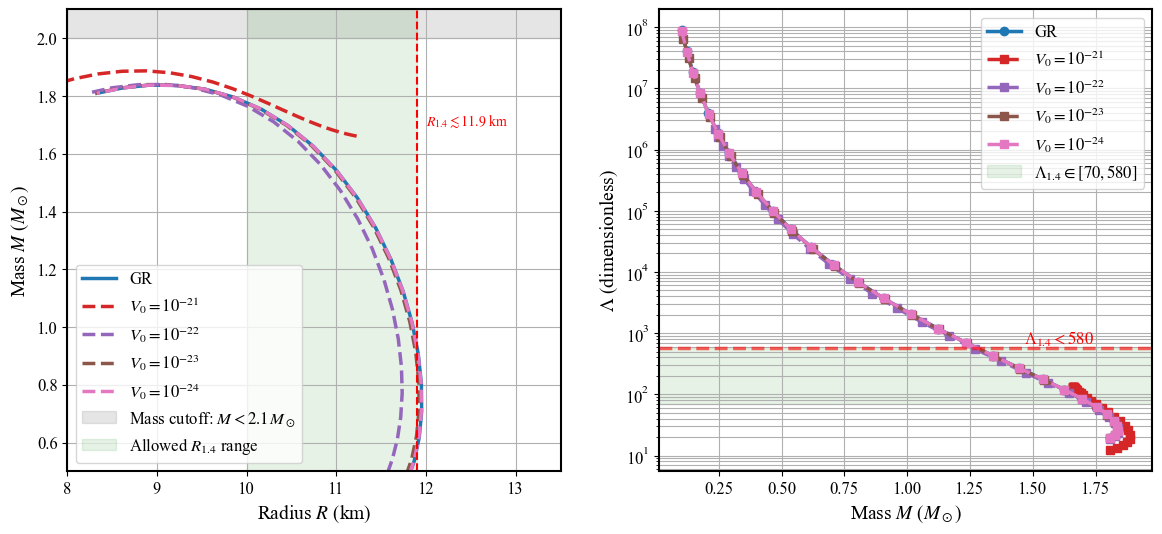}
    \caption{\small Left panel: Mass-radius curves for neutron stars in GR (blue) and scalar-tensor gravity (dashed) with \(\phi(r) \propto r^{1/2}\) and \(V(\phi)=V_0\,\phi^{-2}\). The mass–radius profiles are computed by solving Eq. \eqref{eq:mod_tov} for a polytropic EoS with adiabatic index \(\gamma=2.2\) and polytropic constant \(K_{\text{ST}}=K_{\text{GR}}=800\) with neutron star radius $R_{1.4} \lesssim 11.9\,\text{km}$. The green shaded band marks the allowed radius range for canonical stars, while the gray region indicates the observationally consistent mass window $M \lesssim 2.1\,M_\odot$. Right panel: Tidal deformability $\Lambda$ as a function of stellar mass. The green shaded region marks the LIGO/Virgo bounds on $\Lambda_{1.4}$, the tidal deformability of a $1.4\,M_\odot$ neutron star, constrained to lie within $70 \lesssim \Lambda_{1.4} \lesssim 580$ \cite{abbott2018gw170817}. For \(V_0 = 10^{-24}\, \text{km}^{-2}\) (cosmologically weak potential), deviations from GR are suppressed (\(\Delta\Lambda/\Lambda_{\text{GR}} \lesssim 1\%\)), consistent with LIGO-Virgo bounds \cite{raaijmakers2020constraining}. Stronger scalar potentials (\(V_0 > 10^{-24}\, \text{km}^{-2}\)) yield measurable deviations, enhancing maximum masses while retaining radii within NICER constraints (\(R \geq 10\, \text{km}\))\cite{choudhury2024nicer}.} 
    \label{fig:main}
\end{figure}

Figure \ref{fig:main} and Table \ref{tab:ns_properties} validate the matching where the interior metric functions \(\alpha(R)\), \(\theta(R)\), and \(\phi(R)\) satisfy the exterior analytic solution Eqs.~\eqref{eq:mod_tov} and \eqref{eq:mass_eq} within numerical tolerance (\(\Delta < 0.03\)). For the tidal deformability calculations, we have assumed a constant Love number \(k_2=0.1\), which is a representative value since realistic neutron star models typically predict in the range $0.1\text{--}0.3$ depending on the EoS and compactness \cite{chatziioannou2020neutron}. Notably, for a range of potential \(10^{-21}\leq V_0 <10^{-47}\), (where the upper bound is the observed dark energy density in \(\, \text{km}^{-2}\), the scalar field remains effectively screened, and deviations from GR are suppressed), the field is stable and \(r_{\text{crit}}<1\) adheres to strong screening the scalar field’s impact and the tidal response is suppressed when the potential becomes weaker. A strong screening (deep within the core) can induce an enhanced maximum mass limit relative to GR while still remaining within NICER and LIGO--Virgo constraints. This has a counter-effect when the coupling strength increases.

\begin{table}[htbp]
\scriptsize
\centering
\caption{\small Neutron star configurations for scalar‐field couplings \(\beta = 10^{-3},\,10^{-4},\,10^{-5}\) obtained by numerically solving the modified TOV equations for \(\phi_0 = 10^{-6}\). Each row corresponds to a distinct combination of the polytropic constant $K$ (in geometric units) and adiabatic index $\gamma$. Only configurations with gravitational mass in the range $1.2 \leq M/M_\odot \leq 2.1$, radius $R \leq 11.9\,\text{km}$, and compactness $\mathcal{C} = GM/(Rc^2) \leq 0.44$  are retained.}
\label{tab:combined_filtered}
\begin{tabular}{ccccccccc}
\toprule
\(K\) & \(\gamma\) 
    & \(M_{\max}\) & \(R_{\max}\) & \(\mathcal{C}_{\max}\) 
    & \(M_{1.4}\) & \(R_{1.4}\) & \(\mathcal{C}_{1.4}\) 
    & \(p_c\) (Pa) \\
\midrule

\multicolumn{9}{c}{\(\displaystyle\beta = 10^{-3}\)} \\ 
\cmidrule(l){1-9}
 500  & 2.20 
    & 1.596 & 7.811 & 0.3018 
    & 1.415 & 9.431 & 0.2216 
    & \(2.738\times10^{29}\) \\
 600  & 2.20 
    & 1.722 & 8.684 & 0.2928 
    & 1.374 & 10.622 & 0.1910 
    & \(1.814\times10^{29}\) \\
 600  & 2.30 
    & 1.315 & 6.967 & 0.2789 
    & 1.315 & 6.967 & 0.2789 
    & \(1.878\times10^{29}\) \\
 700  & 2.20 
    & 1.836 & 8.960 & 0.3026 
    & 1.360 & 11.548 & 0.1739 
    & \(2.116\times10^{29}\) \\
 700  & 2.30 
    & 1.416 & 7.187 & 0.2909 
    & 1.392 & 7.420 & 0.2771 
    & \(2.191\times10^{29}\) \\
 800  & 2.30 
    & 1.500 & 7.378 & 0.3004 
    & 1.417 & 8.104 & 0.2582 
    & \(2.503\times10^{29}\) \\
 900  & 2.30 
    & 1.574 & 7.549 & 0.3079 
    & 1.404 & 8.784 & 0.2360 
    & \(2.816\times10^{29}\) \\
1000 & 2.30 
    & 1.638 & 7.704 & 0.3141 
    & 1.432 & 9.235 & 0.2290 
    & \(3.129\times10^{29}\) \\
1100 & 2.30 
    & 1.700 & 8.114 & 0.3095 
    & 1.358 & 9.880 & 0.2030 
    & \(2.524\times10^{29}\) \\
1200 & 2.30 
    & 1.757 & 8.251 & 0.3145 
    & 1.353 & 10.307 & 0.1939 
    & \(2.753\times10^{29}\) \\
1300 & 2.30 
    & 1.813 & 8.666 & 0.3090 
    & 1.443 & 10.542 & 0.2022 
    & \(2.186\times10^{29}\) \\
1400 & 2.30 
    & 1.865 & 8.790 & 0.3133 
    & 1.422 & 10.956 & 0.1917 
    & \(2.355\times10^{29}\) \\
1500 & 2.30 
    & 1.915 & 9.212 & 0.3071 
    & 1.390 & 11.356 & 0.1808 
    & \(1.850\times10^{29}\) \\

\midrule
\multicolumn{9}{c}{\(\displaystyle\beta = 10^{-4}\)} \\
\cmidrule(l){1-9}
 500  & 2.20 
    & 1.596 & 7.800 & 0.3022 
    & 1.416 & 9.416 & 0.2222 
    & \(2.738\times10^{29}\) \\
 600  & 2.20 
    & 1.722 & 8.671 & 0.2933 
    & 1.376 & 10.602 & 0.1917 
    & \(1.814\times10^{29}\) \\
 600  & 2.30 
    & 1.316 & 6.960 & 0.2792 
    & 1.316 & 6.960 & 0.2792 
    & \(1.878\times10^{29}\) \\
 700  & 2.20 
    & 1.836 & 9.259 & 0.2928 
    & 1.363 & 11.524 & 0.1747 
    & \(1.573\times10^{29}\) \\
 700  & 2.30 
    & 1.416 & 7.180 & 0.2913 
    & 1.393 & 7.412 & 0.2775 
    & \(2.191\times10^{29}\) \\
 800  & 2.30 
    & 1.501 & 7.370 & 0.3007 
    & 1.417 & 8.094 & 0.2586 
    & \(2.503\times10^{29}\) \\
 900  & 2.30 
    & 1.574 & 7.540 & 0.3083 
    & 1.405 & 8.773 & 0.2365 
    & \(2.816\times10^{29}\) \\
1000 & 2.30 
    & 1.638 & 7.695 & 0.3144 
    & 1.433 & 9.223 & 0.2295 
    & \(3.129\times10^{29}\) \\
1100 & 2.30 
    & 1.700 & 8.104 & 0.3099 
    & 1.360 & 9.865 & 0.2035 
    & \(2.524\times10^{29}\) \\
1200 & 2.30 
    & 1.757 & 8.241 & 0.3149 
    & 1.355 & 10.291 & 0.1945 
    & \(2.753\times10^{29}\) \\
1300 & 2.30 
    & 1.813 & 8.655 & 0.3094 
    & 1.445 & 10.526 & 0.2028 
    & \(2.186\times10^{29}\) \\
1400 & 2.30 
    & 1.865 & 8.779 & 0.3138 
    & 1.425 & 10.939 & 0.1924 
    & \(2.355\times10^{29}\) \\
1500 & 2.30 
    & 1.915 & 9.199 & 0.3075 
    & 1.393 & 11.336 & 0.1815 
    & \(1.850\times10^{29}\) \\

\midrule
\multicolumn{9}{c}{\(\displaystyle\beta = 10^{-5}\)} \\
\cmidrule(l){1-9}
 500  & 2.20 
    & 1.632 & 7.627 & 0.3160 
    & 1.402 & 9.630 & 0.2150 
    & \(1.512\times10^{29}\) \\
 500  & 2.30 
    & 1.244 & 6.251 & 0.2940 
    & 1.244 & 6.251 & 0.2940 
    & \(1.565\times10^{29}\) \\
 600  & 2.20 
    & 1.765 & 8.411 & 0.3100 
    & 1.514 & 11.097 & 0.2015 
    & \(1.002\times10^{29}\) \\
 600  & 2.30 
    & 1.356 & 6.478 & 0.3091 
    & 1.356 & 6.478 & 0.3091 
    & \(1.878\times10^{29}\) \\
 700  & 2.20 
    & 1.889 & 8.937 & 0.3121 
    & 1.679 & 11.616 & 0.2135 
    & \(8.684\times10^{28}\) \\
 700  & 2.30 
    & 1.448 & 6.667 & 0.3207 
    & 1.399 & 7.271 & 0.2841 
    & \(2.191\times10^{29}\) \\
 800  & 2.30 
    & 1.525 & 6.828 & 0.3300 
    & 1.397 & 8.036 & 0.2568 
    & \(2.503\times10^{29}\) \\
 900  & 2.30 
    & 1.599 & 7.199 & 0.3280 
    & 1.415 & 8.602 & 0.2430 
    & \(2.065\times10^{29}\) \\
1000 & 2.30 
    & 1.667 & 7.572 & 0.3251 
    & 1.391 & 9.280 & 0.2214 
    & \(1.682\times10^{29}\) \\
1100 & 2.30 
    & 1.730 & 7.947 & 0.3215 
    & 1.396 & 9.901 & 0.2083 
    & \(1.356\times10^{29}\) \\
1200 & 2.30 
    & 1.792 & 8.073 & 0.3278 
    & 1.416 & 10.727 & 0.1950 
    & \(1.480\times10^{29}\) \\
1300 & 2.30 
    & 1.849 & 8.448 & 0.3233 
    & 1.493 & 11.030 & 0.2000 
    & \(1.175\times10^{29}\) \\
1400 & 2.30 
    & 1.905 & 8.562 & 0.3287 
    & 1.567 & 11.316 & 0.2046 
    & \(1.266\times10^{29}\) \\
1500 & 2.30 
    & 1.958 & 8.940 & 0.3235 
    & 1.639 & 11.587 & 0.2090 
    & \(9.940\times10^{28}\) \\
\bottomrule
\end{tabular}
\label{tab:ns_properties}
\end{table}

The scalar-induced anisotropy $\Delta p$ modifies the pressure gradient within the neutron star, effectively stiffening the EoS in the high-density core region ($r < r_{\text{crit}}$) while leaving the outer layers ($r > r_{\text{crit}}$) approximately GR-like. For the following computations, we scale the scalar potential strength as $V_0 \propto \phi_0^{4}$, and adopt a relativistic polytropic EoS with $\gamma = 2.2$ and $K = 800$ (in geometric units). Figure \ref{fig:compact} shows a log-log plot of compactness versus scalar amplitude $\phi_0$ for several values of the scalar coupling $\beta$, following the screening-motivated parameter space outlined in \cite{Sagunski2017Neutron}. As expected, smaller values of $\phi_0$ reduce the unscreened scalar region, leading to lower compactness, while intermediate amplitudes enhance central pressure support through stronger scalar gradients. The field does not strongly modify the exterior structure, such as the radius or tidal deformability, which would happen if $r_{\rm crit}$ were close to or outside the surface. This behavior confirms the screening transition radius $r(0) < r_{\text{crit}} < R_s$ remains deep in the core, where high densities suppress the scalar field’s effective range $m_{\text{eff}}^{-1}$ i.e., a direct signature that the transition region lies well within the star, where the density is high. 

For potential strengths below $V_0 \lesssim 10^{-24}\,\text{km}^{-2}$, the scalar field’s influence on the tidal deformability becomes negligible, yielding $\Delta \Lambda / \Lambda_{\text{GR}} \lesssim 1\%$, which remains well within the limits inferred from GW170817 and GW190425 \cite{Abbott2017GW170817:, abbott2020gw190425}. We analyze the resulting stellar configurations for scalar couplings $\beta \in \{10^{-2}, 10^{-3}, 10^{-4}, 10^{-5}\}$. For $\beta = 10^{-2}$, the model yields a maximum mass of $M_{\text{max}} = 1.837\,M_\odot$ and radius $R = 11.827\,\text{km}$ at $\phi_0 = 10^{-3}$, which is below the $\sim 2.0\,M_\odot$ lower bound required by NICER observations. Hence, large couplings are observationally disfavored due to insufficient central pressure. However, for smaller couplings $\beta = 10^{-3}, 10^{-4},$ and $10^{-5}$, the maximum mass is almost constant $M_{\text{max}} = 2.425\,M_\odot$ with radius $R_{\rm max} = 7.775\,\text{km}$, corresponding to increasingly small scalar amplitudes ($\phi_0 \approx 2.15 \times 10^{-4}$ to $2.15 \times 10^{-6}$). This represents a significant $\sim 12.27\%$ increase in maximum mass compared to GR predictions (\(M_{\rm max}^{\rm GR} \approx 2.16 \, M\odot\)) with the same EoS without any scalar field. The resulting scalar configurations are likely to have suppressed tidal deformability ${\Lambda_{\rm ST}/}{\Lambda_{\rm GR}} = 0.137$ , comfortably within LIGO–Virgo bounds. These findings suggest that scalar couplings $\beta \lesssim 10^{-3}$ are both astrophysically viable and potentially favored, while larger couplings are ruled out under current constraints.

\begin{figure}
    \centering
    \includegraphics[width=0.65\linewidth]{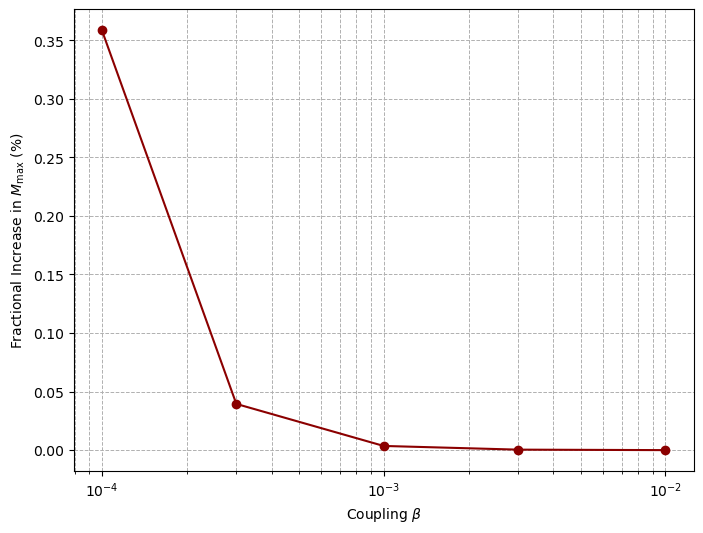}
    \caption{\small Fractional increase in the maximum neutron star mass $M_{\max}$ as a function of scalar–matter coupling $\beta$, for a fixed scalar potential $V_0 = 10^{-26}\,\text{km}^{-2}$. We fixed the base scalar amplitude at \(\phi_0 = 10^{-6}\), and $K_{ST} = 800$. At low $\beta$, scalar gradients enhance the pressure support, yielding a modest increase by 0.36\% in $M_{\max}$ compared to GR. However, for $\beta \gtrsim 10^{-3}$, strong anisotropic stresses from steep scalar gradients suppress this effect, leading to near-GR or sub-GR mass limits. As \(\beta\) increases from \(10^{-4}\) to \(10^{-3}\), mass enhancement drops sharply to below 0.05\%, and near zero. This signals the turnover point where the anisotropic pressure gradient starts to counteract its stabilizing influence.}
    \label{fig:fractional}
\end{figure}

\section{Anisotropy near Compact Objects}

Scalar field configurations near compact objects must satisfy stringent stability criteria so that equilibrium solutions remain physically meaningful even in the presence of strong gravitational gradients \cite{mustafa2020physically}. Beyond a critical mass, such systems may transition to gravitational collapse, ultimately forming a black hole. Following our discussion from previous sections, here, we examine how the gradient of the scalar field and its effective potential generate anisotropic fluid profiles, thereby inducing deviations in the radial and tangential pressures. In this context, the anisotropy, which arises naturally from non-minimal couplings such as \(f(\phi)R\), contributes to exotic gravitational configurations and distinguishes these systems from standard compact stars that are typically characterized by negative principal pressures \cite{horvat2013dark}.

\begin{figure}
    \centering
    \includegraphics[width=\linewidth]{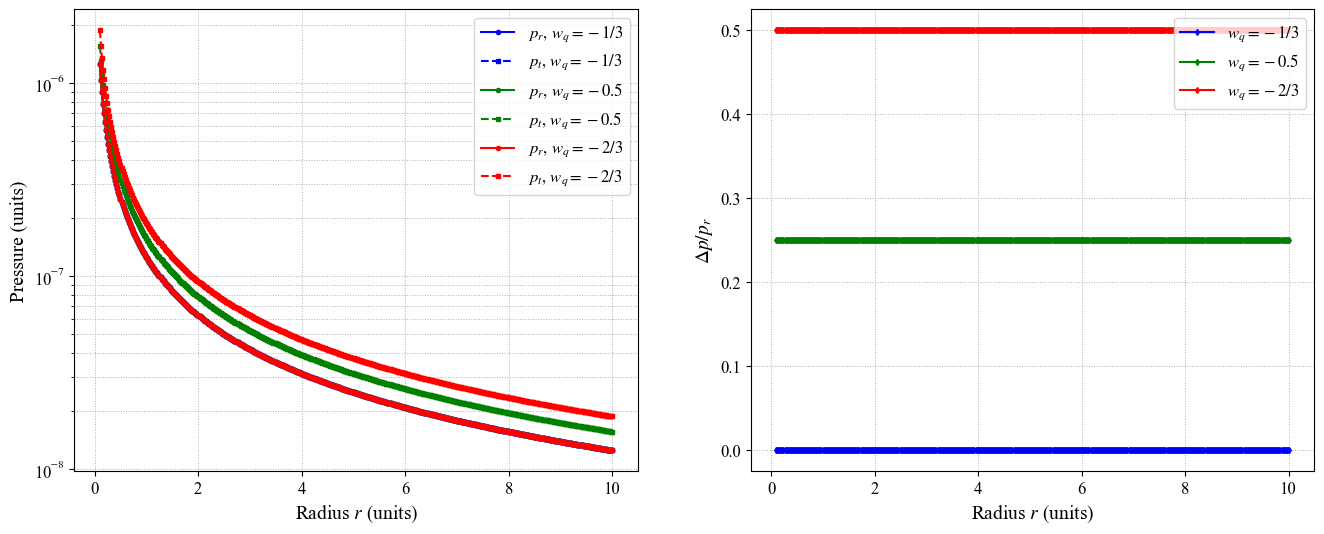}
    \caption{\small Radial profiles of pressure components vs. Normalized anisotropy profile: In the left panel, for \(w_q=-1/3\) (blue curves), \(f(\phi)=0\) causes it to coincide with green, while for \(w_q=-0.5\) and \(w_q=-2/3\) (green and red curves, respectively) a clear difference emerges. The right panel shows the normalized anisotropy ratio \(\Delta p/p_r\) versus radius, which quantifies the relative deviation between \(p_t\) and \(p_r\). Anisotropy is most pronounced in the inner regions, where the scalar field gradient is steep and decreases with increasing radius.}
    \label{fig:anisotropy}
\end{figure}

The energy density \(\rho_\phi\) and the pressure components of the scalar field are derived from its spatial gradient and effective potential and are given by:
\begin{equation}
    \rho_\phi = \frac{1}{2}\left(\frac{d\phi}{dr}\right)^2 + V_{\text{eff}}(\phi), \quad p_r = \frac{1}{2}\left(\frac{d\phi}{dr}\right)^2 - V_{\text{eff}}(\phi), \label{eq:density1}
\end{equation}
where \(p_r\) represents the radial pressure. Here, the anisotropy parameter relates to the EoS parameter \(w_q\) via \(\Delta p = {-1}/{2}(3w_q + 1)\rho_\phi.\) For quintessence-like anisotropy (\(w_q \in (-1, -1/3)\)), \(\Delta p > 0\), indicating \(p_t > p_r\). This distinguishes scalar-induced anisotropy from perfect fluids (\(\Delta p = 0\)). To quantify such deviations from isotropy, we define the tangential pressure as:
\begin{equation}
    p_t = \frac{1}{2}\left(\frac{d\phi}{dr}\right)^2 - V_{\text{eff}}(\phi) + \Delta p, \label{eq:density2}
\end{equation}
with the anisotropy parameter \(\Delta p = p_t - p_r\) capturing the difference induced by the scalar field gradients. The scalar field pressure \(p_\phi\) is then defined as the average pressure in spherical symmetry:
\begin{equation}
    p_\phi = \frac{1}{3}\left(p_r + 2p_t\right),
\end{equation}
which, upon substitution of Eqs.~\eqref{eq:density1} and \eqref{eq:density2}, simplifies to:
\begin{equation}
    p_\phi = \frac{1}{2}\left(\frac{d\phi}{dr}\right)^2 - V_{\text{eff}}(\phi) + \frac{2}{3}\Delta p.
\end{equation}
Thus, in the isotropic limit (\(\Delta p = 0\)), the scalar pressure reduces to that of a perfect fluid scalar configuration. Figure \ref{fig:anisotropy} illustrates the radial profiles of the pressure components induced by the scalar field.

The effective potential in Eq.~(\ref{eq:effectpoten}) combines the intrinsic scalar potential with the kinetic coupling term. In the limit \(V(\phi) \to 0\) (perfect fluid), the effective potential \( V_{\text{eff}}(\phi) \) is dominated by the scalar field’s kinetic coupling and matter interactions:
\begin{equation}
    V_{\text{eff}}(\phi) = -\frac{1}{2r^2}\mathcal{K}_{\phi,\phi} - \sum_i \frac{\beta_i \rho_i(r)}{M_{\text{pl}}},
\end{equation}
where the \(-1/2\) factor is from the Taylor expansion of \(\mathcal{K}(\phi)\)  to second order, and the negative sign arises from the stress-energy tensor’s conventions in Einstein’s equations. Near a compact object (at small \(r\)), the dominant contribution to the scalar field gradient in Eq. \eqref{eq:fullequ} can be approximated as:
\begin{equation}
    \frac{d\phi}{dr} \approx -\left[ \left(m_\phi^2 + \sum_i \frac{3\beta_i^2 \rho_i(r)}{M_{\text{pl}}^2}\right)\phi + \frac{\mathcal{K}_{\phi,\phi}}{r^2} + \sum_i \frac{\beta_i \rho_i(r)}{M_{\text{pl}}} \right]. \label{eq:dominantcom}
\end{equation}
Quantum corrections to light scalar field theories, such as the chameleon field, impose constraints on the mass such that, for reliable predictions of fifth forces, these corrections must remain minimal \cite{khoury2013chameleon}. In practice, this requirement leads to an upper bound \(m_\phi < 0.0073\,\text{eV}\) and a lower bound \(m_\phi > 0.0042\,\text{eV}\). These bounds arise from fifth-force constraints \cite{Upadhye2012Quantum}, where quantum corrections to the chameleon potential \(V(\phi)\) become significant if \(m_\phi\) exceeds the inverse interparticle spacing in dense matter (\(\sim 0.01\,\text{eV}\)). Ignoring this small-scale fluctuation arising from quantum corrections, Eq.~(\ref{eq:dominantcom}) can be simplified to:
\begin{equation}
    \frac{d\phi}{dr} \sim -\left[ m_\phi^2 \phi + \frac{\mathcal{K}_{\phi,\phi}}{r^2} + \sum_i \frac{\beta_i \rho_i(r)}{M_{\text{pl}}} \right]. \label{eq:loopcorr}
\end{equation}
In general, the effective mass may vary within the compact object due to a non-trivial density profile, but it is important to note that the above system of equations is incomplete without loop corrections that account for pressure fluctuations and shear from quantum interactions. In regions where the bound is violated, our linearized approach must be re-evaluated, and a fully nonlinear quantum field theoretic treatment might be warranted. We defer a detailed exploration of these loop corrections to future work.

By substituting Eq.~(\ref{eq:loopcorr}) into Eqs.~\eqref{eq:density1} and \eqref{eq:density2}, we obtain:
\begin{equation}
    \rho_\phi = \frac{1}{2}\left[ m_\phi^2 \phi + \frac{\mathcal{K}_{\phi,\phi}}{r^2} + \sum_i \frac{\beta_i \rho_i(r)}{M_{\text{pl}}} \right]^2 - \frac{1}{2r^2}\mathcal{K}_{\phi,\phi} - \sum_i \frac{\beta_i \rho_i(r)}{M_{\text{pl}}}, \label{eq:rhophi}
\end{equation}
\begin{equation}
    p_r = \frac{1}{2}\left[ m_\phi^2 \phi + \frac{\mathcal{K}_{\phi,\phi}}{r^2} + \sum_i \frac{\beta_i \rho_i(r)}{M_{\text{pl}}} \right]^2 + \frac{1}{2r^2}\mathcal{K}_{\phi,\phi} + \sum_i \frac{\beta_i \rho_i(r)}{M_{\text{pl}}}. \label{eq:pr}
\end{equation}
Similarly, the tangential pressure \(p_t\) is given by Eq.~\eqref{eq:density2}. In spherical symmetry, the conservation law reduces to the continuity equation, and for physically acceptable models, both \(p_r\) and \(p_t\) must remain positive. For a static configuration (with \(\partial \rho_\phi/\partial t = 0\)), the combined evolution of the fluid density and radial pressure obeys:
\begin{equation}
    \frac{1}{r^2}\frac{d}{dr}\left[r^2 (\rho_\phi + p_r)\right] \geq 0, \label{eq:relation}
\end{equation}
where the sum \(\rho_\phi + p_r\) can be expressed as:
\begin{equation}
    \rho_\phi + p_r = \left[ m_\phi^2 \phi + \frac{\mathcal{K}_{\phi,\phi}}{r^2} + \sum_i \frac{\beta_i \rho_i(r)}{M_{\text{pl}}} \right]^2. \label{eq:densitypress}
\end{equation} At small radii, any significant anisotropic stress could induce a strong repulsive effect, leading to pronounced deviations between the radial and tangential pressures. As \(r\) increases, \(\mathcal{K}_{\phi,\phi}\) diminishes, and the configuration gradually approaches isotropy (\(\Delta p \to 0\)); since the scalar field becomes more uniformly distributed, and its coupling with external densities is reduced. 

\begin{figure}[h]
    \centering
    \includegraphics[width=0.65\linewidth]{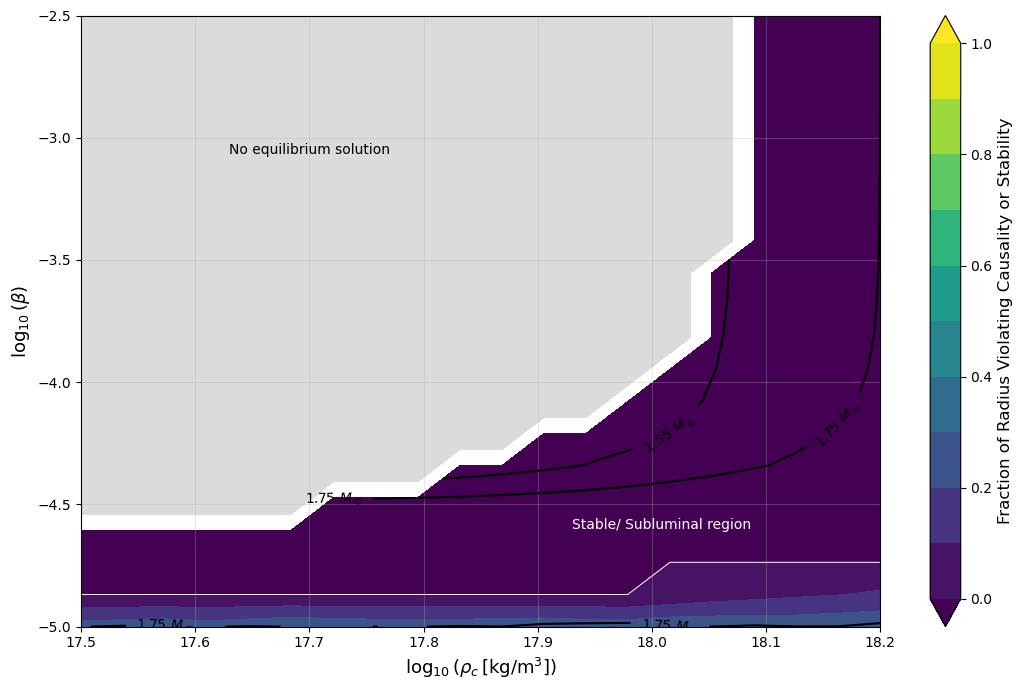}
    \caption{Contour plot showing the fraction of the stellar radius that violates either the causality condition $(v_s^2 < 1)$ or the relativistic stability condition $(\Gamma > 4/3)$, as a function of the central density $\rho_c$ and the scalar coupling parameter $\beta$, both plotted on logarithmic scales. Dark blue regions indicate stars that are entirely stable and subluminal, satisfying both physical constraints. The light gray shaded area marks configurations where no equilibrium solution exists, likely due to divergence in scalar field gradients or strong violations of the hydrostatic balance. Intermediate colored regions (blue to green) indicate partial violations of the stability or causality conditions across the stellar interior. Overlaid black contours represent constant total gravitational mass levels of $1.55\,M_\odot$ and $1.75\,M_\odot$, corresponding to scalarized solutions consistent with the mass bounds in Figure \ref{fig:energycon}.} 
    \label{fig:contourplot}
\end{figure}

While Eq.~\eqref{eq:densitypress} ensure that \(\rho_\phi + p_r \geq 0\), the weak energy condition (WEC) for the combined quantity, \(\rho_\phi + p_\phi\) imposes \(V_{\text{eff}}(\phi) \leq 0\) and \(\Delta p > 0\). These conditions constrain the types of viable scalar potentials in canonical neutron star models. At small radii, the dominance of \(\Delta p\) can lead to a divergence, which may ultimately cause a collapse into a black hole, in accordance with Buchdahl's theorem \cite{maurya2018generalized}. Here, \(V_{\text{eff}} \leq 0\) indicates that the kinetic energy \((\partial_r\phi)^2\) dominates over potential contributions, preventing ghost instabilities. Otherwise, it can lead to exotic configurations—often associated with negative tension—that deviate from standard GR predictions by producing tighter orbits and enhanced precession effects. In regions of strong gravitational gradients, satisfying the WEC may require \(\dot{\phi} \geq 0\), a condition that can be met by appropriately approximating \(V(\phi)\) and \(\dot{\phi}\) near equilibrium, yielding density and pressure profiles analogous to those of a continuous scalar fluid \cite{koivisto2015scalar, thirukkanesh2012exact, noureen2019models, mardan2018new, singh2020anisotropic}. To distinguish equilibrium configurations from pathological ones, we also compute the spatial fraction of the star that violates either the causality or adiabatic stability condition, as shown in Fig.\ref{fig:contourplot}. Stable equilibrium stars—particularly those near $1.4\,M_\odot$ and the maximum-mass threshold—occupy a narrow window in $(\rho_c, \beta)$ parameter space that delineates the boundary between equilibrium and non-equilibrium regimes.

\begin{figure}[h]
    \centering
    \includegraphics[width=0.65\linewidth]{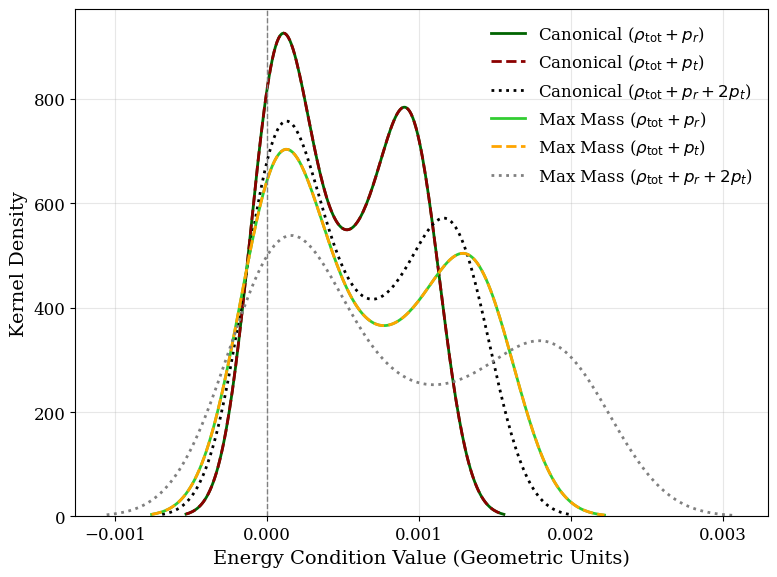}
    \caption{\small Kernel density estimates (KDE) of energy condition terms for canonical ($M \approx 1.4\,M_\odot$) and maximum-mass scalarized neutron star configurations. Each curve represents the distribution of energy condition values across the radial profile, including total energy density and anisotropic pressures induced by the scalar field. The weak energy condition (WEC) is evaluated using both radial ($\rho_{\rm tot} + p_r$) and tangential ($\rho_{\rm tot} + p_t$) components, while the strong energy condition (SEC) is given by $\rho_{\rm tot} + p_r + 2p_t$. Uncertainty estimates, assuming a 5\% variation in central density $\rho_c$, yield: for the canonical star, $M_{\rm 1.4} = 1.554 \pm 0.375\,M_\odot$, $R_{\rm 1.4} = 11.95 \pm 0.90\,\text{km}$; and for the maximum-mass configuration, $M_{\rm max} = 1.750 \pm 0.280\,M_\odot$, $R_{\rm max} = 11.35 \pm 1.11\,\text{km}$. All curves lie predominantly in the positive domain, indicating that both WEC and SEC are satisfied throughout the interior for the chosen parameters: $V_0 = 10^{-24} \,\mathrm{km}^{-2}$, $\phi_0 = 10^{-6}$, and polytropic index $\gamma = 2.2$. The vertical dashed line at zero marks the threshold for energy condition violation.}
    \label{fig:energycon}
\end{figure}

Compared to other anisotropic models \cite{maurya2018generalized, sharif2020anisotropic, morales2018charged, raposo2019anisotropic, maurya2021anisotropic, singh2020anisotropic, mak2003anisotropic, Maurya2018Anisotropic}, Eqs. \ref{eq:rhophi}–\ref{eq:pr} ensures finite, regularized energy density as \( r \to 0 \), eliminating coordinate singularities by enforcing \( \mathcal{K}_{\phi,\phi} \to 0 \). The scalar field \( \phi(r \to 0) \) must remain nonzero and finite at the centre, ensuring positive contributions with negligible quantum corrections, requiring \( \rho_\phi(0) > 0 \) for stable central fluid configurations (see Appendix B). If \( \phi(r \to 0) \) grows large near compact objects, the potential’s quadratic term (\( m_\phi^2 \phi(r) \)) dominates over linear contributions while maintaining a positive central energy density given that \( \phi(r) \) and \( \rho_m(r) \) remain finite and balanced. However, mismatches between radial and tangential stresses in anisotropic systems induce cracking instabilities \cite{herrera1992cracking, abreu2007sound}, controlled by sound-speed conditions. We have verified that the scalar-fluid configuration jointly satisfies both the Weak and Strong Energy Conditions (SEC) pointwise throughout the stellar interior using the total energy density $\rho_{\rm tot} = \rho (r) + \rho_\phi$  where $\rho(r)$ is the baryonic matter density extracted from the polytropic EoS. These profiles, evaluated numerically for both the canonical and maximum-mass configurations, are shown in Figure \ref{fig:energycon}.  All solutions that pass causality and stability tests (summarized in Table \ref{tab:stability-scan}) are also found to satisfy WEC and SEC. In particular, the scalar energy density $\rho_q \propto r^{-1}$ remains dominant only in the outer stellar layers and falls off quickly near the surface, where its influence is screened; consistent with the Kiselev-type fluid interpretation introduced in Section 1.

\section{Future Prospects}
Compact objects such as neutron and boson stars do not have event horizons and possess finite radii with complex internal structures. However, their external gravitational fields may approximate a Kiselev-like solution at distances \(r \gg R\), where internal density variations become negligible and the spacetime is governed by the anisotropic fluid’s effective energy density \(\rho_q(r) \propto r^{-3(w_q+1)}\). This approximation holds because, at distances much greater than the object's radius \(r>r_{\text{crit}}\), any fine-grained internal density variations become negligible, and the spacetime is effectively determined by the surrounding energy density \cite{tsizh2014distribution}. Interestingly, such compact configurations can exceed the Buchdahl limit \cite{PhysRev.116.1027, andreasson2008sharp}, classifying them as Kiselev ClePhOs (Clean-Photon Stars) or other types of compact star-like objects \cite{pani2018gravitational}. Investigating whether these configurations can exist across a wide range of masses presents an open possibility for addressing challenges related to black hole mimickers as their compactness depends on the balance between scalar screening (\(\mathcal{K}_{\phi,\phi} \propto (\partial_r\phi)^2\)) and dark energy dominance (\(w_q \in (-1, -1/3)\)) across a broad mass range (see Sec. 4–5). While most black hole mimickers tend to approach black hole compactness within a narrow mass range, ClePhOs and similar compact objects demonstrate greater flexibility \cite{shapiro2024black}. This characteristic makes them valuable candidates for exploring gravitational signatures and potential astrophysical deviations from classical black hole models. 

In this paper, we examined a scalar field that does not carry a monopolar scalar charge in the exterior spacetime. This ensures that fifth-force constraints are satisfied asymptotically, although interior scalar gradients generate significant local anisotropy. The presence of scalar charge influences the deflection angle in both weak and strong gravitational lensing scenarios \cite{Eiroa2014Strong}. In strong lensing, this leads to distinct observational signatures when compared to Schwarzschild and Reissner-Nordström black holes \cite{Pang2018Gravitational}. While theories such as Galileon and DHOST predict notable deviations in gravitational lensing potential, others, like Brans-Dicke's theory, suggest more constrained effects. For example, Brans-Dicke scalar fields have shown limited impact on strong lensing observables, indicating that the coupling parameter may be tightly constrained in terms of its influence on lensing phenomena \cite{Zhang2017Strong, Sarkar2006Strong}. However, measuring the effects caused by scalar field gradients may require unprecedented sensitivity with observations. The next-generation Event Horizon Telescope (ngEHT) angular resolution, which is sub-microarcseconds (approximately 0.1 $\mu$as), will allow it to resolve azimuthal distortions in the photon rings around supermassive black holes (SMBHs) caused by scalar-field gradients \cite{tiede2022measuring}. 

While our analysis has focused on static, compact configurations, the underlying scalar-tensor framework carries potential implications for late-time cosmology. In the Kiselev limit, spatial gradients are bounded by metric contributions to deviations in radial and tangential pressure components. Without these bounds,  the scalar field is unable to generate anisotropic stresses, leading to an isotropic configuration that complies with the Weak Energy Condition.  In this context, backreaction is dynamically introduced into the scalar field's effective potential \(V_{\text{eff}}(\phi)\), which self-consistently adjusts the metric's stress-energy components. Furthermore, quasi-local scalar gradients near compact objects can induce environment-dependent screening of the luminosity distance-redshift relation through disformal coupling.  If any disformal redshift corrections systematically bias luminosity distances at the (\( \sim 10^{-5} - 10^{-4} \)) level, they may offer a partial resolution to the observed Hubble tension. This results in non-kinematic contributions to redshift and may preferentially bias late-time measurements of the Hubble constant \(H_0\) (from types of supernovae like SNe Ia or Cepheids), while leaving early-universe constraints (such as CMB sound horizons) unaffected. While these cosmological implications are promising, detecting scalar-mediated effects in compact objects requires unprecedented instrumental precision.  A viable test would involve cross-correlating low-redshift SN Ia luminosity residuals with sky regions containing compact objects capable of hosting unscreened scalar profiles. While speculative, future work could test this by cross-correlating DESI's \cite{karim2025desi} \( w(z) \) with compact object observables (such as lensing asymmetries  \(\Delta\theta_{\text{asym}} \propto \mathcal{K}_\phi \) at \( r > r_{\text{crit}} \) \cite{tutusaus2023first}) to isolate environment-dependent scalar effects.

\section{Discussion}

Scalar-tensor theories often suggest that scalarization can generate anisotropic pressures within compact stars, commonly seen in Bowers-Liang neutron stars or Vaidya-inspired fluid parameterizations. In contrast, we investigate a scalar dark energy model that generates anisotropy through scalar field gradients (\(\partial_r \phi\)) by introducing non-minimal couplings and Horndeski-inspired kinetic terms. These gradients are modulated to source anisotropic pressures (\(\Delta p \)) while evading no-hair theorems through curvature-dependent screening. Importantly, the chameleon mechanism dynamically suppresses fifth forces in high-density regimes (\(\rho \gtrsim 10^{15}\,\mathrm{g/cm^3}\)). This reconciles local repulsive anisotropies (\(\Delta p \propto r^{-1}\)) with solar system constraints while allowing for unscreened quintessence-like behavior (\(w_q \in (-1, -1/3)\)) to dominate cosmologically—a hybrid role not well explored in the literature. We studied a simple scalar field profile with power law ansatz \(\phi(r) \propto r^{-(3w_q + 1)/2}\), following Kiselev's parameterization of dark energy. In the special case of quintessence with \(w_q = -{2}/{3}\), this reduces to \(\phi(r) \sim r^{1/2}\). However, the ansatz is valid in regimes where the gravitational curvature is not extremely high—that is, in the outer regions of compact stars or in regimes where \(\phi\) remains sufficiently small such that \(\beta \phi/M_{\rm pl} \ll 1\). In the innermost regions of neutron stars, where densities and curvature are extremely high, non–linear effects may render this approximation less accurate. In such cases, a full numerical solution by solving the modified TOV equation is preferred.

Unlike phenomenological Kiselev models, which impose anisotropy through a static parameter \(w_q\), our model suggests that dark energy’s influence is not uniformly distributed but is concentrated in regions of strong gravitational effects, such as the core/crust of a neutron star. These regions are typically described by piecewise EoS models that match nuclear theory with crustal physics. In contrast, our approach employs a single relativistic polytropic EoS, wherein the physical effects of the crust are absorbed through the tuning of the polytropic parameters.  Crucially, we introduce a dynamically motivated critical radius (\(r_{\rm crit}\)), below which the scalar field is effectively screened due to the high effective mass arising from dense matter coupling. While this screening occurs deep within the stellar core, it does not contradict the consistency of the nuclear EoS, as the scalar field functions as a perturbative sector layered on top of the existing matter content rather than replacing it entirely. Beyond \(r_{\rm crit}\), scalar gradients dominate over potential terms (\(V(\phi) \ll (\partial_r\phi)^2\)), leading to significant deviations in radial and tangential pressures.  This stiffens the EoS and modifies neutron star compactness (\(\mathcal{C} =M/R\)) in a way that compact objects act as \say{transition zones} where the scalar’s effective mass (\(m_\phi\)) grows sharply with matter density, suppressing long-range forces. Although the radial gradients of scalar field \((\partial_r\phi)^2\) generate measurable deviations from GR in the compact object structure, anisotropic stresses in canonical neutron stars generally follow NICER fit (\(\mathcal{C}_{\rm 1.4} \lesssim 0.22 \, \, \text{and} \,\,  \mathcal{C}_{\rm max} \lesssim 0.36 \)). However, the radii of such configurations (\( R_{\rm 1.4} \lesssim 10 \, \text{km} \)) remains within the conservative bound of NICER and LIGO-Virgo constraints (see Table \ref{tab:ns_properties}), which favor \( R_{\rm s} \geq 10 \, \text{km} \) for \( M \sim 1.4 M_\odot \). From this standpoint, any deviation \(>5\%\) from GR remains inconsistent with current constraints from LIGO/Virgo and NICER, which tightly constrain the coupling constant \(\beta\) in our model.

Numerical integration of the modified TOV equations (Eq. \eqref{eq:mod_tov}) using our fiducial parameters (\(\gamma = 2.2\), \(K = 800\)) yields neutron star observables compatible with multi-messenger constraints. The maximum-mass scalarized configuration achieves $M_{\text{max}} \approx 1.75 \pm 0.280\,M_\odot$ with radius $R \approx 11.35 \pm 1.11 \,\mathrm{km}$, while the canonical configuration with  $M_{\rm 1.4} = 1.554 \pm 0.375\,M_\odot$, attains $R_{\rm 1.4} = 11.95 \pm 0.90\,\text{km}$. The corresponding tidal deformabilities lie within the LIGO–Virgo band (\(70 \leq \Lambda_{1.4} \leq 580\)), and the compactness \( \mathcal{C} \sim 0.1\)–\(0.2 \) is within the NICER and LIGO/Virgo constraints. These configurations are causal, stable, and satisfy energy conditions throughout the star. It should be noted that $M_{\text{max}} = 1.75\,M_\odot$ corresponds to a configuration where the scalar field amplitude $\phi_0 = 10^{-6}$ is fixed at a cosmologically relevant value. This choice prioritizes consistency with large-scale scalar field dynamics, potentially linked to dark energy phenomenology. However, as shown in Table \ref{tab:stability-scan}, we have also explored an extended parameter space where the scalar amplitude is allowed to vary. In those configurations, the suppression of scalar field effects in high-density regions is enhanced, enabling equilibrium solutions with maximum masses exceeding $2\,M_\odot$, in agreement with the NICER constraint from PSR J0740+6620. Therefore, while the cosmologically motivated scenario leads to lower maximum masses, the scalarized configurations are flexible enough to accommodate observed high-mass pulsars under adjusted field strengths and couplings.

The scalar field’s effective mass remains confined to a narrow window bounded from below by gravitational wave dispersion constraints (\(m_\phi \gtrsim 10^{-22}\,\text{eV}\)) and from above by fifth-force experiments and quantum stability bounds (\(m_\phi \lesssim 10^{-2}\,\text{eV}\)). These results, validated across a range of scalar configurations (Figure \ref{fig:compact}), show that the maximum mass remains almost the same across weak couplings when scalar contribution perturbs neutron star observables by less than \(5\%\) compared to GR due to scalar-induced pressure support. The observed decrease in the maximum neutron star mass for \(\beta = 10^{-2}\) by approximately \(24\%\) (compared to lower values of \(\beta\) ) indicates that strong scalar coupling suppresses central pressure support, softens the EoS, and reduces gravitational stability. In our calibrated models (\(\gamma = 2.2\), \(K = 800\)), the maximum mass decreases from \(1.76\,M_\odot\) in the GR limit to \(1.75\,M_\odot\) for \(\beta = 10^{-3}\), even as the radius contracts slightly. While this shift is modest, it places upper limits on the coupling parameter \(\beta \lesssim 10^{-3}\) to remain consistent with observed high-mass neutron stars. This inverse trend between \(\beta\) and \(M_{\text{max}}\) provides a potential signature for constraining scalar-tensor theories i.e, increasing \(\beta\) enhances the scalar-induced pressure support via the scalar energy density \(\rho_q \propto r^{-1}\), which dominates over the outward anisotropic pressure support from \(\Delta p = (\partial_r \phi)^2\). This imbalance increases the gravitational binding energy, producing marginally more compact stars and suppressing the maximum mass. Although anisotropic stresses generally stiffen the EoS, here, the net effect of the scalar field acts as an inward pull due to its energy density, counteracting pressure support. This is expected since the chameleon screening mechanism likely suppresses scalar effects in the dense core (\(\rho \gtrsim 10^{15} \, \mathrm{g/cm^3}\)), while unscreened regions in the outer layers soften the pressure gradient, effectively increasing the binding energy and reducing the pressure support for massive field configurations.

While our analysis demonstrates the scalar field's compatibility with a modified potential at short scales, it is important to recognize the model's limitations. The model’s current formulation assumes a static, spherically symmetric scalar field configuration with a power-law ansatz, limiting its applicability to highly asymmetric systems such as merging neutron stars or rapidly rotating compact objects, where time-dependent scalar gradients and frame-dragging effects may dominate. The reliance on quasi-static approximations for the scalar field’s effective mass neglects transient phenomena and backreaction from spacetime curvature fluctuations, potentially underestimating instabilities in rapidly evolving systems. This necessitates Bayesian calibration against multi-messenger data to mitigate degeneracies between scalar couplings (\(\beta_i\)), potential strengths (\(V_0\)), and EoS uncertainties. Furthermore, the exclusion of quantum corrections beyond fifth-force bounds (\(m_\phi \lesssim 0.01\,\text{eV}\)) restricts predictive power in ultra-dense regimes (\(\rho \gtrsim 10^{16}\,\text{g/cm}^3\)), where vacuum polarization and loop effects may alter the scalar potential’s behavior. While our calibrated polytropic EoS captures the qualitative behavior of nuclear matter, including realistic crust–core microphysics via tabulated EoS would allow for more accurate modeling of the critical radius. 

\section{Acknowledgments}
The authors would like to express their gratitude to Christ (Deemed to be University) for funding this research under the Seed Money Project. VJ and KA thank the Inter-University Centre for Astronomy and
Astrophysics (IUCAA), Pune, India, for the Visiting Associateship.
\section{ORCID}
\vspace{2mm}
Pradosh Keshav MV: https://orcid.org/0000-0001-7104-3102
\vspace{2mm} \newline
Jithesh V: https://orcid.org/0000-0002-6449-9643
\vspace{2mm} \newline
Kenath Arun:  https://orcid.org/0000-0002-2183-9425
\section{Declaration of competing interest}
\begin{enumerate}
    \item Author Contributions: Pradosh Keshav M.V. conceived the core idea, developed the theoretical framework, and wrote the manuscript. Jithesh V. and Kenath Arun contributed to the theoretical analysis, numerical implementation, and interpretation of the physical results.
    \item Funding: Seed money scheme, Sanction No. CU-ORS-SM-24/29.
    \item Data Availability Statement: Not applicable.
    \item Conflicts of Interest: The authors declare no conflict of interest.
\end{enumerate}
\section{Declaration of Generative AI and AI-assisted technologies in the writing process}
During the preparation of this work, the author, Pradosh Keshav, used Grammarly to correct grammar mistakes and make some paragraphs more structured. After using this tool/service, the author reviewed and edited the content as needed and take(s) full responsibility for the content of the publication.

\appendix
\section*{Appendix A}
\renewcommand{\theequation}{A\arabic{equation}}
\label{sec:appendixA}
Here, we derive  Eq.~\eqref{eq:derivativeoffr} assuming that the scalar-metric coupling is dependent on the non-minimal coupling function \(f(\phi)\) and the metric ansatz for the Kiselev spacetime. We begin with the scalar-tensor action in Eq.~\eqref{eq:gen action}, specializing to a static, spherically symmetric geometry described by the line element:
\begin{equation}
    ds^2 = -f(r)\,dt^2 + \frac{dr^2}{f(r)} + r^2 d\Omega^2, \label{eq:A1}
\end{equation}
where \(f(r) = 1 - 2M/r - a/r^{3w_q + 1}\) and \(a\) is a normalization constant that quantifies the magnitude of the anisotropic fluid’s contribution to the spacetime geometry. While \(a\) appears as a phenomenological parameter in the original Kiselev metric, its value is dynamically fixed by the scalar field’s boundary conditions, as will be demonstrated below.

For a scalar field profile \(\phi(r)\), the modified Einstein equations take the form:
\begin{equation}
    f(\phi) G_{\mu\nu} + \nabla_\mu \nabla_\nu f(\phi) - g_{\mu\nu} \Box f(\phi) = 8\pi T_{\mu\nu}^{(\phi)}, \label{eq:A2}
\end{equation}
with \(T_{\mu\nu}^{(\phi)}\) given by Eq.~\eqref{eq:emtensor}. The \((t,t)\) and \((r,r)\) components of Eq.~\eqref{eq:A2} are:
\begin{equation}
    G_{tt} = \frac{f(r)(1 - f(r))}{r^2} - \frac{f'(r)}{r}, \quad G_{rr} = -\frac{1 - f(r)}{r^2 f(r)} - \frac{f'(r)}{r f(r)}. \label{eq:ttandrr} 
\end{equation} By substituting Eq. \eqref{eq:A2} into Eq. \eqref{eq:ttandrr} and equating it to \(8\pi T^{(\phi)}_{\mu\nu}\), we obtain the following relation:
\begin{equation}
    f(\phi)\left(\frac{f'(r)}{r} + \frac{f(r) - 1}{r^2}\right) + \frac{f''(\phi)(\phi')^2 + f'(\phi)\phi''}{2} = 8\pi\rho, \label{eq:A3} 
\end{equation}
where primes ($'$) denote radial derivatives.  It is important to note that consistency with the scaling of the anisotropic fluid’s density, \(\rho \propto r^{-3(w_q + 1)}\), requires a power-law ansatz of the form \(\phi(r) = \phi_0 r^{-(3w_q + 1)/2}\), with the scalar amplitude \(\phi_0\) defined at a reference length. Varying the action in Eq. \eqref{eq:gen action} with respect to \(\phi\), we get:
\begin{equation}
    \nabla^2 \phi = \frac{1}{\omega(\phi)}\left[V'(\phi) - \frac{f'(\phi) R}{16\pi}\right], \label{eq:A4} 
\end{equation}
By matching terms order-by-order in \(r\), we find that the non-minimal coupling must adopt an exponential form:
\begin{equation}
    f(\phi) = \exp\left(\frac{3\beta_i\phi}{M_{\rm pl}}\right), \label{eq:A5} \
\end{equation}
to cancel the divergent terms arising from the scalar potential \(V(\phi) \propto \phi^{-2}\). 

The metric function \(f(r)\) implicitly depends on \(\phi(r)\) through the integration constant \(a\), which is determined by boundary conditions at spatial infinity. As \(r\) approaches infinity (\(r \to \infty\)), the scalar field behaves as \(\phi \to \phi_0 r^{-(3w_q + 1)/2}\), while the metric simplifies to \(f(r) \to 1 - \frac{2M}{r}\). By matching the subleading term \(\mathcal{O}(r^{-(3w_q + 1)})\), we can express \(a\) follows:
\begin{equation}
    a = \frac{8\pi \phi_0^2 (3w_q + 1)}{3(1 - w_q)} \exp\left(\frac{3\beta_i\phi_0 r^{-(3w_q + 1)/2}}{M_{\rm pl}}\right), \label{eq:A6} 
\end{equation}
where \(\phi_0\) sets the scalar amplitude. This shows that \(a\), though seemingly a normalization constant in the Kiselev metric, is fundamentally tied to the scalar field’s amplitude (\(\phi_0\)) and coupling (\(\beta_i\)) (with dimension \(L^{3w_q +1}\) ).  The exponential dependence of \(a\) remains constant for fixed values of \(\phi_0\) and \(\beta_i\), even as it dynamically adjusts in response to the spatial gradients of the scalar field. 

Taking the derivative of \(f(r)\) with respect to \(\phi\) yields:
\begin{equation}
    \frac{\partial f(r)}{\partial \phi} = -\frac{1}{r^{3w_q + 1}} \frac{\partial a}{\partial \phi} = -\frac{3\beta_i}{M_{\rm pl}} \frac{a}{r^{3w_q + 1}} \exp\left(\frac{3\beta_i\phi}{M_{\rm pl}}\right), \label{eq:A7} 
\end{equation}
which recovers Eq.~\eqref{eq:derivativeoffr} from the main text. The consistency of this derivation with the Einstein field equations can be verified by direct substitution of Eqs.~\eqref{eq:A5}--\eqref{eq:A7} into Eq.~\eqref{eq:A3}, confirming that the scalar’s backreaction dynamically adjusts \(f(r)\) to maintain equilibrium. By substituting \(f(\phi)\), \(\phi(r)\), and \(a\) into Eq. \eqref{eq:A3}, we arrive at \(\rho \propto r^{-3(w_q + 1)}\), thereby validating the ansatz.

\section*{Appendix B}
\renewcommand{\theequation}{B\arabic{equation}}
In this appendix, we outline the procedure for assessing the stability of a scalar field configuration discussed in the main text, as it interacts with a compact object. In this context, we analyze the eigenvalues of the Jacobian matrix associated with the linearized version of the modified Klein–Gordon equation, in order to establish conditions for stability.

Starting from the scalar field equation of motion presented in the main text:
\begin{equation}
    \Box \phi = V_{\phi}(\phi) + \frac{1}{2r^2} \frac{\partial \dot{\phi}_\infty^2(r)}{\partial \phi} + \sum_i \frac{\beta_i}{M_{\text{pl}}} \rho_i \exp\!\left(\frac{3\beta_i \phi}{M_{\text{pl}}}\right), \label{eq:A12}
\end{equation}
where \(V_{\phi}(\phi) \equiv \frac{dV}{d\phi}\) and \(\rho_i\) denotes the local matter density. By assuming spherical symmetry, the d'Alembertian reduces to:
\begin{equation}
    \nabla^2 \phi = \frac{d^2 \phi}{dr^2} + \frac{2}{r} \frac{d\phi}{dr}.
\end{equation}
Thus, Eq.~\eqref{eq:A12} becomes:
\begin{equation}
    \frac{d^2 \phi}{dr^2} + \frac{2}{r} \frac{d\phi}{dr} = V_{\phi}(\phi) + \frac{1}{2r^2} \frac{\partial \dot{\phi}_\infty^2(r)}{\partial \phi} + \sum_i \frac{\beta_i}{M_{\text{pl}}} \rho_i \exp\!\left(\frac{3\beta_i \phi}{M_{\text{pl}}}\right).  \label{eq:A13}
\end{equation}
To convert this second-order differential equation into a first-order autonomous system, we define the auxiliary variables \(x = \frac{d\phi}{dr}, \quad y = \phi.\)
The system then takes the following form:
\begin{align}
    \frac{dx}{dr} &= -\frac{2}{r}\,x + V_{\phi}(y) + \frac{1}{2r^2}\frac{\partial \dot{\phi}_\infty^2(r)}{\partial y} + \sum_i \frac{\beta_i}{M_{\text{pl}}} \rho_i \exp\!\left(\frac{3\beta_i y}{M_{\text{pl}}}\right),\label{eq:A14} \\
    \frac{dy}{dr} &= x. \label{eq:A15}
\end{align}

To determine the equilibrium points \((x_{\text{eq}}, y_{\text{eq}})\) of the system, we set the following equation:
\begin{equation}
    x_{\text{eq}} = 0, \quad V_{\phi}(y_{\text{eq}}) + \frac{1}{2r^2}\frac{\partial \dot{\phi}_\infty^2(r)}{\partial y} + \sum_i \frac{\beta_i}{M_{\text{pl}}} \rho_i \exp\!\left(\frac{3\beta_i y_{\text{eq}}}{M_{\text{pl}}}\right) = 0.  \label{eq:A16}
\end{equation}To assess the stability of these equilibrium points, we linearize the system by introducing small perturbations, \(x \to x_{\text{eq}} + \delta x\) and \(y \to y_{\text{eq}} + \delta y\). The linearized system can then be written in matrix form as follows:
\begin{equation}
    \frac{d}{dr} \begin{bmatrix} \delta x \\ \delta y \end{bmatrix} = J \begin{bmatrix} \delta x \\ \delta y \end{bmatrix},  \label{eq:A17}
\end{equation}
where the Jacobian matrix \(J\) is defined as:
\begin{equation}
    J = \begin{bmatrix}
    -\frac{2}{r} & V_{\phi\phi}(y_{\text{eq}}) + \frac{\partial^2}{\partial y^2}\left(\frac{1}{2r^2}\dot{\phi}_\infty^2(r)\right) + \sum_i \frac{\beta_i^2}{M_{\text{pl}}^2} \rho_i \exp\!\left(\frac{3\beta_i y_{\text{eq}}}{M_{\text{pl}}}\right) \\
    1 & 0
    \end{bmatrix}.  \label{eq:A18}
\end{equation}

The eigenvalues \(\lambda\) of \(J\) must satisfy the characteristic equation:
\begin{equation}
    \lambda^2 + \frac{2}{r}\lambda - \left[ V_{\phi\phi}(y_{\text{eq}}) + \frac{\partial^2}{\partial y^2}\left(\frac{1}{2r^2}\dot{\phi}_\infty^2(r)\right) + \sum_i \frac{\beta_i^2}{M_{\text{pl}}^2} \rho_i \exp\!\left(\frac{3\beta_i y_{\text{eq}}}{M_{\text{pl}}}\right) \right] = 0. \label{eq:A19}
\end{equation}Since the discriminant of this quadratic equation is given by:
\begin{equation}
    \Delta = \left(\frac{2}{r}\right)^2 - 4\left[ V_{\phi\phi}(y_{\text{eq}}) + \frac{\partial^2}{\partial y^2}\left(\frac{1}{2r^2}\dot{\phi}_\infty^2(r)\right) + \sum_i \frac{\beta_i^2}{M_{\text{pl}}^2} \rho_i \exp\!\left(\frac{3\beta_i y_{\text{eq}}}{M_{\text{pl}}}\right) \right],  \label{eq:A20}
\end{equation}
with eigenvalues:
\begin{align}
    \lambda &= -\frac{1}{r} \pm \sqrt{-\Delta}, \quad \text{if } \Delta < 0, \label{eq:A21a} \\
    \lambda &= -\frac{1}{r} \pm \sqrt{\Delta}, \quad \text{if } \Delta > 0. \label{eq:A21b}
\end{align}For the system to be stable, the effective mass squared of the scalar field must be positive, i.e.,
\begin{equation}
    m_\phi^2 = V_{\phi\phi}(y_{\text{eq}}) + \frac{\partial^2}{\partial y^2}\left(\frac{1}{2r^2}\dot{\phi}_\infty^2(r)\right) + \sum_i \frac{\beta_i^2}{M_{\text{pl}}^2} \rho_i \exp\!\left(\frac{3\beta_i y_{\text{eq}}}{M_{\text{pl}}}\right) > 0.  \label{eq:A21c}
\end{equation}

When \(\Delta < 0\), the eigenvalues are complex with negative real parts (i.e., \(-\frac{1}{r}\)), indicating oscillatory stability or a stable spiral—characteristic of chameleon screening in dense environments. Conversely, if \(\Delta > 0\) and all eigenvalues are real and negative, the system is a stable node with exponential decay dominating, as expected in low-density, quintessence-like regimes. Near compact objects, where the matter density \(\rho_i\) dominates, \(m_\phi^2\) is typically positive, and the scalar field remains undetectable \(m_{\phi}\gg H_0\) .  Conversely, in low-density environments, the dominance of the potential gradient \(V_{\phi}(\phi)\) and the kinetic term \(\dot{\phi}_\infty^2(r)\) might alter stability conditions, with  \(m_{\phi}\ll H_0\) driving cosmic acceleration. We have summarised the above analysis in Tab. \ref{tab:stability} for a comparison between different regimes.

\begin{table}[ht]
\centering
\scriptsize
\begin{tabularx}{\textwidth}{|p{1.6cm}|p{1.5cm}|p{1.5cm}|p{1.6cm}|p{1.9cm}|p{2cm}|}
\hline
\textbf{Environment} & \textbf{Energy Density (\(\rho_i\))} & \textbf{Dominant Contribution} & \textbf{Effective Mass (\(m_\phi^2\))} & \textbf{Discriminant (\(\Delta\))} & \textbf{Stability Behaviour} \\ \hline
\textbf{Near Compact Objects} & High (\(\rho_i \gg V_{\phi\phi}(y)\)) & Matter coupling term & Positive due to dominance of \(\rho_i\). & Negative (\(\Delta < 0\)) & Stable spiral (oscillatory damping). \\ \hline
\textbf{Far from Compact Objects} & Low (\(\rho_i \ll V_{\phi\phi}(y)\)) & Scalar potential term & Depends on the curvature of \(V(\phi)\). & Can be positive (\(\Delta > 0\)) or negative (\(\Delta < 0\)) & Stable node (real damping) or stable spiral. \\ \hline
\textbf{Transition Region} & Intermediate (\(\rho_i \sim V_{\phi\phi}(y)\)) & Combination of matter and potential contributions & Can vary depending on specific model parameters (e.g., \(\beta_i\), \(V(\phi)\), and \(\dot{\phi}_\infty^2\)). & Transition between \(\Delta < 0\) and \(\Delta > 0\). & Stability depends on effective mass sign. \\ \hline
\textbf{Very Low-Density Regions} & Negligible (\(\rho_i \approx 0\)) & Potential term \(V_{\phi\phi}(y)\) & Determined entirely by \(V_{\phi\phi}(y)\): stable if \(V_{\phi\phi}(y) > 0\); unstable if \(V_{\phi\phi}(y) < 0\). & Depends on \(V_{\phi\phi}(y)\): stable or unstable. & Stability relies on curvature of \(V(\phi)\). \\ \hline
\end{tabularx}
\caption{\small Stability analysis of the scalar field based on energy density near and far from the compact objects based on a static scalar field configuration from Eq. \eqref{eq:A21c}.}
\label{tab:stability}
\end{table}
\appendix
\section*{Appendix C}

\begin{figure}
    \centering

    \subfloat[Convergence of $r_{\rm crit}$ iteration for $\phi_0 = 10^{-3}$, $ m_{\text{eff}} = 10^{-6} \,\text{eV}$, and $\beta = 10^{-3}$.]{
        \includegraphics[width=0.43\textwidth]{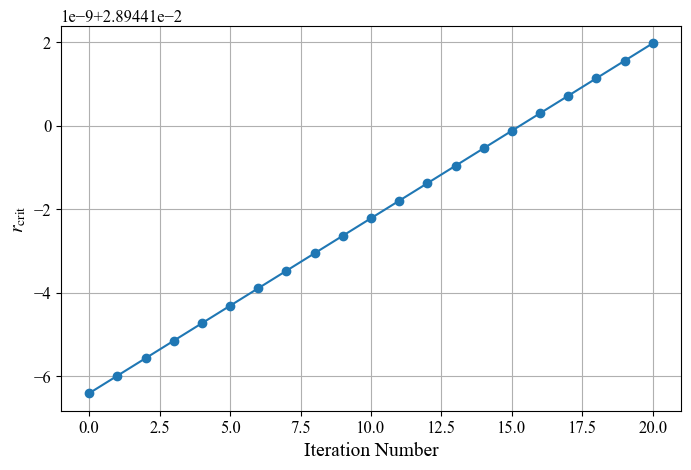}
        \label{fig:sgsh}
    }
    \hfill
    \subfloat[Dependence of $r_{\rm crit}$ on $\beta$ for $\phi_0 = 10^{-3}$ and $ m_{\text{eff}} = 10^{-6} \,\text{eV}$.]{
        \includegraphics[width=0.43\textwidth]{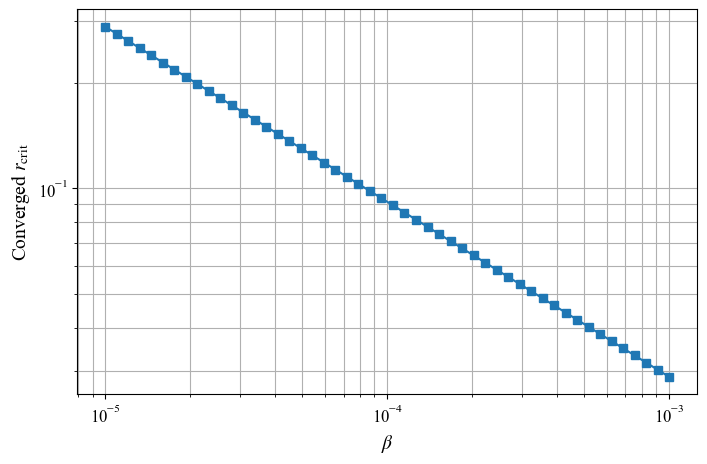}
        \label{fig:sdgd}
    }

    \subfloat[Converged $r_{\rm crit}$ for $\phi_0 = 10^{-3}$,$ m_{\text{eff}} = 10^{-24} \,\text{eV}$, and $\beta = 10^{-3}$.]{
        \includegraphics[width=0.43\textwidth]{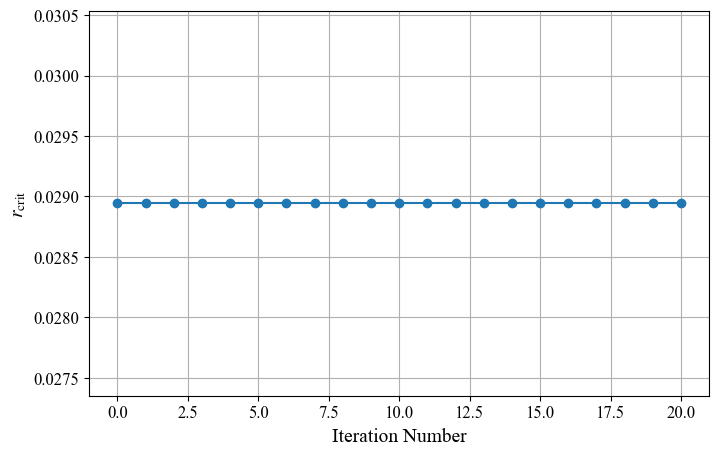}
        \label{fig:dgd}
    }
    \hfill
    \subfloat[Dependence of $r_{\rm crit}$ on $\beta$ for $\phi_0 = 10^{-3}$ and $ m_{\text{eff}} = 10^{-24}\,\text{eV}$.]{
        \includegraphics[width=0.43\textwidth]{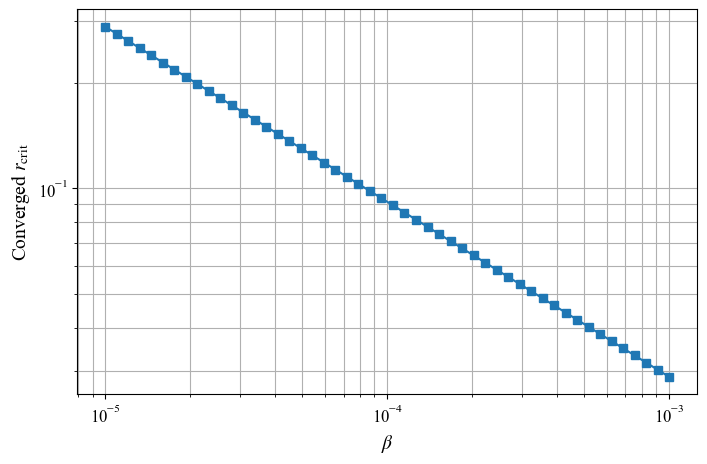}
        \label{fig:dgdj}
    }
    
    \caption{ \small The iterative convergence of \( r_{\text{crit}} \) for a neutron star with a scalar-matter coupling of \( \beta = 10^{-3} \) and a scalar amplitude of \( \phi_0 = 10^{-3} M_{\text{pl}} \) is examined. Despite variations in \( m_{\text{eff}} \), the initial estimate \( r_{\text{crit}}^{(0)} \) remains nearly constant, indicating that the scalar-matter balance is primarily influenced by the parameter \( a \) and the core radius \( r_{\text{core}} \). The initial estimate \( r_{\text{crit}}^{(0)} \approx 0.0289 \, \text{km} \), derived from Eq. \eqref{eq:criticalradius}, evolves through successive iterations that account for the scalar's effective mass \( m_{\text{eff}} \), stabilizing to \( r_{\text{crit}} \approx 0.03 \, \text{km} \) within three cycles.  }
    \label{fig:iteration}
\end{figure}

Here, we detail the iterative procedure used to determine the critical radius \( r_{\rm crit} \) – the radial distance within the compact object where the scalar-field–induced anisotropic effects transition from being dominant (unscreened) to subdominant (screened). As outlined in Section~4 of the main text, the initial estimate is defined by \(r_{\rm crit}^{(0)} = \sqrt{{a\,M_{\rm pl}}/{\beta\,r_{\rm core}}},\) with \( M_{\rm pl} =1 \) in geometric units. In natural units, this scales approximately as \(0.3 \, M_{\rm pl}^{1/2}\), with appropriate dimensional rescaling applied using conversion factors described earlier. Subsequent iterations (illustrated in Figure~\ref{fig:iteration}) refine \(r_{\rm crit}^{(0)}\) by incoperating the scalar backreaction via the effective scalar mass \( m_{\text{eff}} \), which sets the scale for Yukawa-type suppression. 

\begin{figure}
    \centering
    \begin{subfigure}[b]{0.42\textwidth}
        \includegraphics[width=\textwidth]{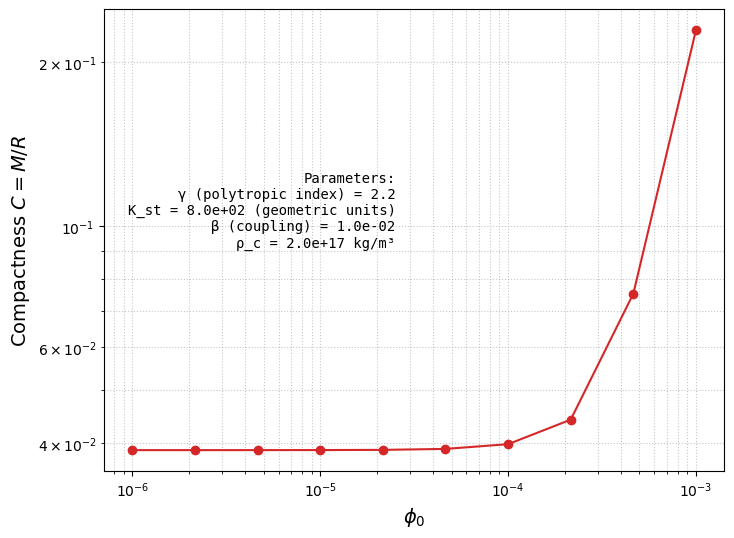}
       
        \label{fig:sub21}
    \end{subfigure}
    \begin{subfigure}[b]{0.46\textwidth}
        \includegraphics[width=\textwidth]{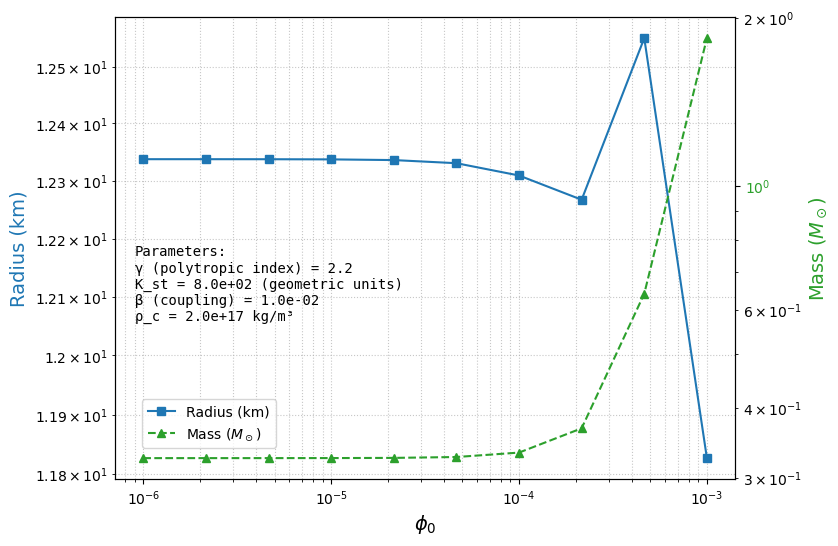}
       
        \label{fig:sub61}
    \end{subfigure}
    \hfill
    \begin{subfigure}[b]{0.42\textwidth}
        \includegraphics[width=\textwidth]{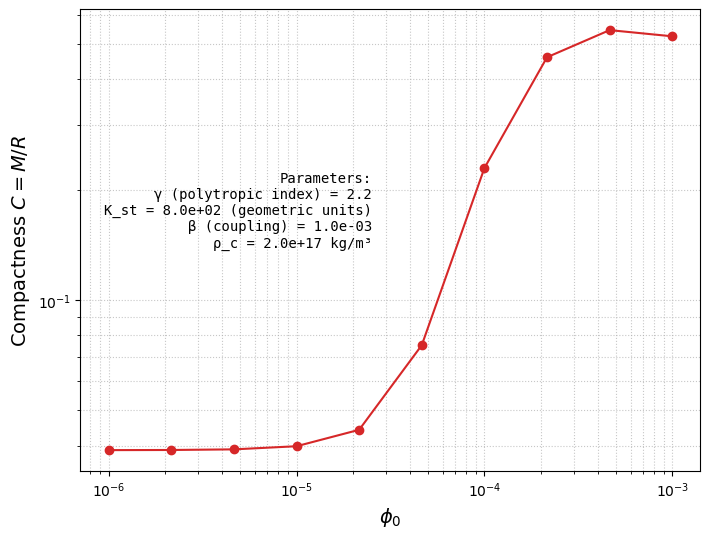}
       
        \label{fig:sub11}
    \end{subfigure}
    \begin{subfigure}[b]{0.44\textwidth}
        \includegraphics[width=\textwidth]{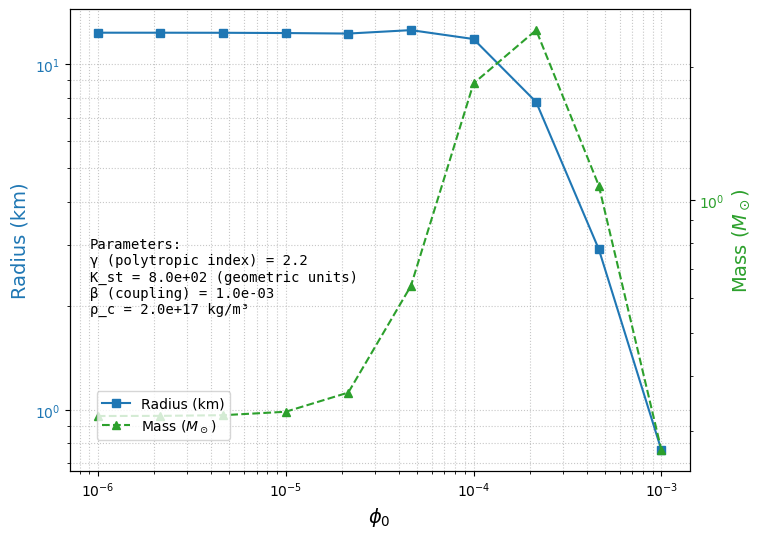}
        
        \label{fig:sub81}
    \end{subfigure}
    \hfill
    \begin{subfigure}[b]{0.42\textwidth}
        \includegraphics[width=\textwidth]{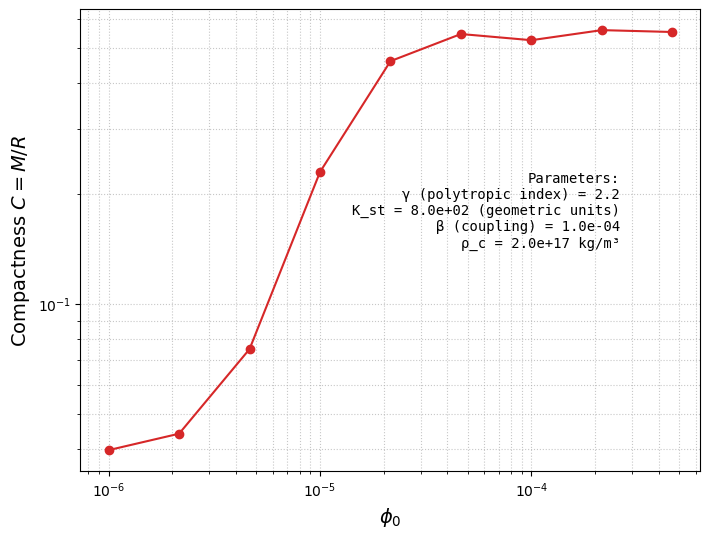}
       
        \label{fig:sub22}
    \end{subfigure}
    \begin{subfigure}[b]{0.44\textwidth}
        \includegraphics[width=\textwidth]{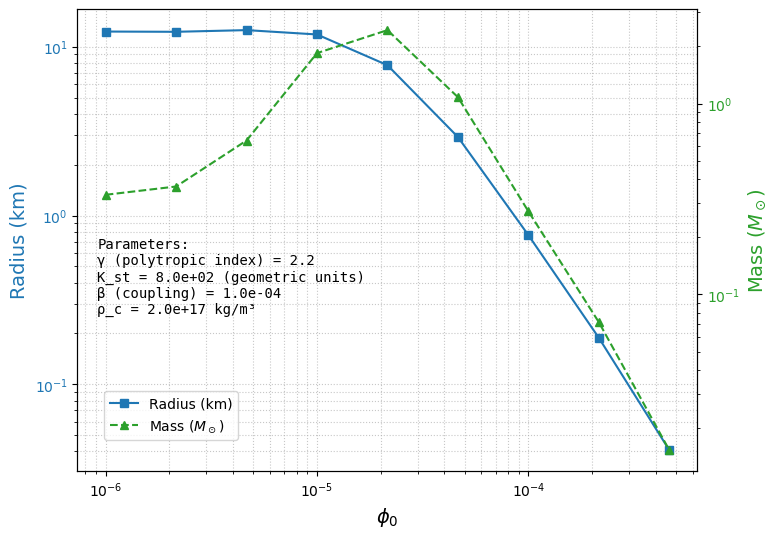}
       
        \label{fig:sub23}
    \end{subfigure}
    \hfill
    \begin{subfigure}[b]{0.42\textwidth}
        \includegraphics[width=\textwidth]{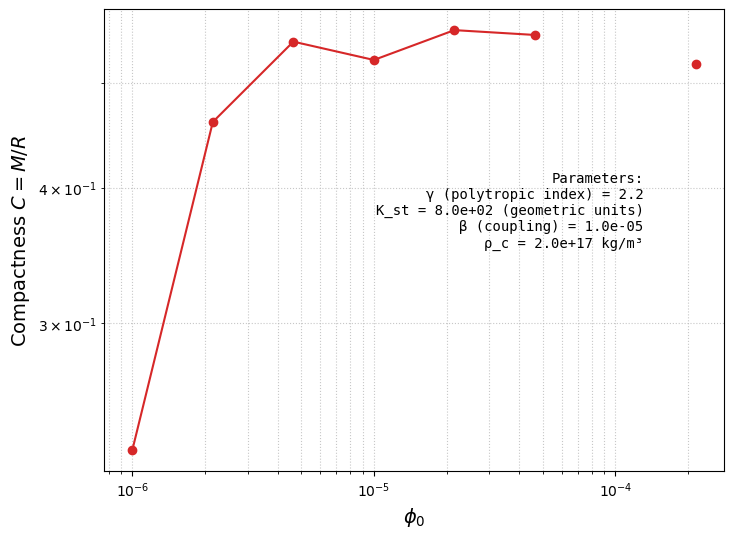}
        
        \label{fig:sub41}
    \end{subfigure}
    \begin{subfigure}[b]{0.44\textwidth}
        \includegraphics[width=\textwidth]{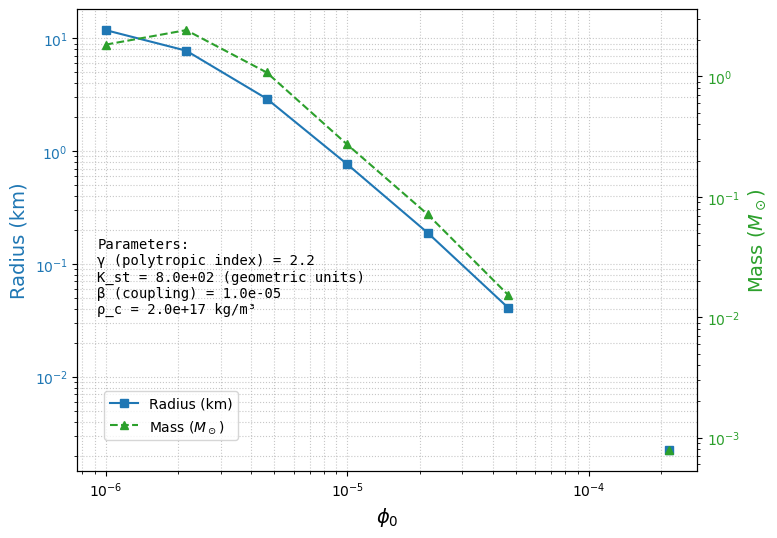}
       
        \label{fig:sub91}
    \end{subfigure}
    \caption{\small  Compactness \(C = M/R\) of neutron stars as a function of the base scalar amplitude \(\phi_0\) (in geometric units) for four different coupling strengths is shown on the left side. The potential scales as \(V_0\propto \phi_0^{4} \) and a decrease in \(\phi_0\) leads to a lower critical (unscreened) radius and, thus, a reduction in compactness.  Stellar radius and mass as functions of the base scalar amplitude \(\phi_0\) for the same set of coupling constants are shown on the right side. These plots illustrate how variations in \(\phi_0\) influence the equilibrium configuration of neutron stars, providing complementary information to the compactness plot.}
    \label{fig:compact}
\end{figure}

We find that the procedure converges within three iterations to a stable solution \( r_{\text{crit}} \sim 0.3 \, \text{km} \) for representative values \( \beta \sim 10^{-3}, \phi_0 \sim 10^{-3} M_{\text{pl}} \). This rapid convergence is a characteristic feature of scalar-tensor systems where screening mechanisms are predominant in high-density environments similar to spontaneous scalarization observed in neutron stars, where iterative methods help resolve nonlinear interactions between scalar \say{hair} and matter. Unlike phase transitions driven by local density thresholds, however, in our case, \( r_{\text{crit}} \) defines a geometric suppression scale tied to the spacetime curvature through the parameter \(a\).

The small magnitude of \( r_{\text{crit}} \) compared to neutron star radii (\( R_{\text{NS}} \sim 10 \, \text{km} \)) confines scalar effects to low-density regions, such as the crust or envelope, while maintaining GR-like behavior in the core. This naturally produces localized anisotropic stresses in the exterior, consistent with the chameleon-like suppression of scalar fields in high-density regions. Figure \ref{fig:compact} visualizes this behavior by mapping the scalar amplitude \( \phi_0 \) against stellar compactness \( \mathcal{C} = M/R \). Even for sub-Planckian amplitudes \( \phi_0 \lesssim 10^{-4} M_{\text{pl}} \) the scalar field can stiffen the outer EoS, increasing the compactness by roughly 1\%  (see Tab. \ref{tab:ns_properties} for those exceeding the Buchdahl limit) compared to GR solutions. The stiffening, while subtle, modifies tidal deformabilities \( \Delta\Lambda/\Lambda_{\text{GR}} \approx -4.5\% \) within LIGO/Virgo bounds \( \tilde{\Lambda} \leq 580 \) at levels detectable by next-generation gravitational wave observatories.

 \begin{table}
 \scriptsize
\centering
\caption{\small Parameter scan over scalar field neutron star configurations: For each triplet of the polytropic index \( \gamma \), coupling constant \( \beta \), and potential amplitude \( V_0 \), we report the resulting stellar mass \( M \) and radius \( R \), along with whether the interior configuration satisfies the causality condition \( v_s^2 < 1 \) and the relativistic adiabatic index stability condition \( \Gamma > 4/3 \).  The table entries correspond to configurations whose central energy densities were scanned over the range \(\rho_c \in [10^{17.5}, 10^{18.2}]\,\text{kg/m}^3\), which covers the typical values for canonical neutron stars with \(\phi_0 = 10\). This choice is motivated by its compatibility with black hole spacetimes and effective chameleon screening near compact sources. Configurations that satisfy both NICER radius and mass constraints are highlighted in bold.}
\begin{tabular}{ccccccccc}
\hline
$\gamma$ & $\beta$ & $V_0$ & $M\,[M_\odot]$ & $R\,[\text{km}]$ & Causal & Stable \\
\hline
1.90 & $1.0\times10^{-4}$ & $1.0\times10^{-24}$ & 5.21 & 25.94 & No  & Yes \\
1.90 & $1.0\times10^{-4}$ & $1.0\times10^{-23}$ & 5.21 & 25.94 & No  & Yes \\
1.90 & $1.0\times10^{-4}$ & $1.0\times10^{-22}$ & 5.21 & 25.94 & No  & Yes \\
1.90 & $1.0\times10^{-4}$ & $1.0\times10^{-21}$ & 5.21 & 25.94 & No  & Yes \\
1.90 & $5.0\times10^{-4}$ & $1.0\times10^{-24}$ & 5.18 & 26.15 & No  & Yes \\
1.90 & $5.0\times10^{-4}$ & $1.0\times10^{-23}$ & 5.18 & 26.15 & No  & Yes \\
1.90 & $5.0\times10^{-4}$ & $1.0\times10^{-22}$ & 5.18 & 26.15 & No  & Yes \\
1.90 & $5.0\times10^{-4}$ & $1.0\times10^{-21}$ & 5.18 & 26.15 & No  & Yes \\
2.00 & $5.0\times10^{-4}$ & $1.0\times10^{-24}$ & 3.77 & 17.80 & No  & Yes \\
2.00 & $5.0\times10^{-4}$ & $1.0\times10^{-23}$ & 3.77 & 17.80 & No  & Yes \\
2.00 & $5.0\times10^{-4}$ & $1.0\times10^{-22}$ & 3.77 & 17.80 & No  & Yes \\
2.00 & $5.0\times10^{-4}$ & $1.0\times10^{-21}$ & 3.77 & 17.80 & No  & Yes \\
2.00 & $1.0\times10^{-3}$ & $1.0\times10^{-24}$ & 3.64 & 17.50 & No  & Yes \\
2.00 & $1.0\times10^{-3}$ & $1.0\times10^{-23}$ & 3.64 & 17.50 & No  & Yes \\
2.00 & $1.0\times10^{-3}$ & $1.0\times10^{-22}$ & 3.64 & 17.50 & No  & Yes \\
2.00 & $1.0\times10^{-3}$ & $1.0\times10^{-21}$ & 3.64 & 17.50 & No  & Yes \\
2.10 & $1.0\times10^{-4}$ & $1.0\times10^{-24}$ & 2.21 & 19.57 & Yes & Yes \\
2.10 & $1.0\times10^{-4}$ & $1.0\times10^{-23}$ & 2.21 & 19.57 & Yes & Yes \\
2.10 & $1.0\times10^{-4}$ & $1.0\times10^{-22}$ & 2.21 & 19.57 & Yes & Yes \\
2.10 & $1.0\times10^{-4}$ & $1.0\times10^{-21}$ & 2.21 & 19.57 & Yes & Yes \\
2.10 & $5.0\times10^{-4}$ & $1.0\times10^{-24}$ & 2.86 & 13.72 & No  & Yes \\
2.10 & $5.0\times10^{-4}$ & $1.0\times10^{-23}$ & 2.86 & 13.72 & No  & Yes \\
2.10 & $5.0\times10^{-4}$ & $1.0\times10^{-22}$ & 2.86 & 13.72 & No  & Yes \\
2.10 & $5.0\times10^{-4}$ & $1.0\times10^{-21}$ & 2.86 & 13.72 & No  & Yes \\
2.10 & $1.0\times10^{-3}$ & $1.0\times10^{-24}$ & 2.75 & 12.70 & No  & Yes \\
2.10 & $1.0\times10^{-3}$ & $1.0\times10^{-23}$ & 2.75 & 12.70 & No  & Yes \\
2.10 & $1.0\times10^{-3}$ & $1.0\times10^{-22}$ & 2.75 & 12.70 & No  & Yes \\
2.10 & $1.0\times10^{-3}$ & $1.0\times10^{-21}$ & 2.75 & 12.70 & No  & Yes \\
\textbf{2.20} & \textbf{$5.0\times10^{-4}$} & \textbf{$1.0\times10^{-24}$} & \textbf{2.06} & \textbf{11.81} & \textbf{Yes} & \textbf{Yes} \\
\textbf{2.20} & \textbf{$5.0\times10^{-4}$} & \textbf{$1.0\times10^{-23}$} & \textbf{2.06} & \textbf{11.81} & \textbf{Yes} & \textbf{Yes} \\
\textbf{2.20} & \textbf{$5.0\times10^{-4}$} & \textbf{$1.0\times10^{-22}$} & \textbf{2.06} & \textbf{11.81} & \textbf{Yes} & \textbf{Yes} \\
\textbf{2.20} & \textbf{$5.0\times10^{-4}$} & \textbf{$1.0\times10^{-21}$} & \textbf{2.06} & \textbf{11.81} & \textbf{Yes} & \textbf{Yes} \\
2.20 & $1.0\times10^{-3}$ &  $1.0\times10^{-24}$ & 2.12 &  9.93 & No  & Yes \\
2.20 & $1.0\times10^{-3}$ &  $1.0\times10^{-23}$ & 2.12 &  9.93 & No  & Yes \\
2.20 & $1.0\times10^{-3}$ &  $1.0\times10^{-22}$ & 2.12 &  9.93 & No  & Yes \\
2.20 & $1.0\times10^{-3}$ &  $1.0\times10^{-21}$ & 2.12 &  9.93 & No  & Yes \\
\hline
\end{tabular}
\label{tab:stability-scan}
\end{table}

To regularize the divergence in $d\phi/dr \sim r^{-1/2}$ at the center, we introduce a small cutoff at $r_{\rm crit}$, below which the scalar field is effectively turned off. This regularized core simulates the stiff behavior of high-density nuclear matter governed by realistic EoS such as SLy or APR. Physically, this corresponds to the assumption that scalar field effects become significant only beyond a critical density, consistent with the screening behavior of chameleon-type fields. Table \ref{tab:stability-scan} summarizes the results of our stability and causality analysis. For each parameter set, we identify the stellar solution closest to the canonical mass $M = 1.4\,M_\odot$ and verify whether the causality condition $v_s^2 < 1$ and the relativistic stability criterion $\Gamma(r) > 4/3$ are satisfied throughout the interior. The cutoff ensures that scalar modifications remain inactive near the center and prevents unphysical divergences in the scalar pressure. This strategy also helps us satisfy the Weak and Null Energy Conditions (see Section 7), even for the maximum-mass configurations.

To obtain stable and causal solutions across the full range of scalar-tensor parameters, we require a higher scalar amplitude ($\phi_0 \sim 10$), while energy conditions (shown in Figure \ref{fig:energycon}) remain satisfied even for lower amplitudes ($\phi_0 \sim 10^{-6}$). Consequently, different diagnostic plots are generated with parameter sets appropriate to the specific physical criterion under investigation. The elevated values of $\phi_0$ required to match neutron star observables do not necessarily invalidate the model. Instead, they emphasize that the relationship between the cosmological \(\phi_0\) and its effective value within dense compact objects is nontrivial and requires context-specific renormalization. Future work may eliminate the need for a sharp cutoff by either employing smooth scalar profiles such as $\phi(r) \propto \tanh(\sqrt{r})$ or by matching the interior to a realistic nuclear EoS. This approach would also allow for comprehensive  MCMC sampling of the interior field profile without singular behaviour.

\bibliographystyle{ws-ijmpa}
\bibliography{sample}

\begin{thebibliography}{100}
\expandafter\ifx\csname urlstyle\endcsname\relax
  \providecommand{\doi}[1]{doi:\discretionary{}{}{}#1}\else
  \providecommand{\doi}{doi:\discretionary{}{}{}\begingroup \urlstyle{rm}\Url}\fi

\bibitem{perlmutter1999measurements}
S.~Perlmutter {\em et~al.}, {\em Astrophysical Journal} {\bf 517}, 565  (1999).

\bibitem{riess1998observational}
A.~G. Riess {\em et~al.}, {\em Astronomical Journal} {\bf 116}, 1009  (1998).

\bibitem{planck2020cosmological}
P.~Collaboration, {\em Astronomy \& Astrophysics}   (2020).

\bibitem{Yang2020Dynamical}
W.~Yang, E.~D. Valentino, S.~Pan, Y.~Wu and J.~Lu, {\em Monthly Notices of the Royal Astronomical Society}   (2020), \doi{10.1093/mnras/staa3914}.

\bibitem{Huterer2001Weak}
D.~Huterer, {\em Physical Review D} {\bf 65},   063001  (2001), \doi{10.1103/PhysRevD.65.063001}.

\bibitem{arun2018alternate}
K.~Arun, S.~Gudennavar, A.~Prasad and C.~Sivaram, {\em Advances in Space Research} {\bf 61}, 567  (2018).

\bibitem{del2017small}
A.~Del~Popolo and M.~Le~Delliou, {\em Galaxies} {\bf 5},  ~17  (2017).

\bibitem{mavromatos2022lambda}
N.~E. Mavromatos, { Lambda-cdm model and small-scale-cosmology “crisis”: from astrophysical explanations to new fundamental physics models}, in {\em The Fifteenth Marcel Grossmann Meeting: On Recent Developments in Theoretical and Experimental General Relativity, Astrophysics, and Relativistic Field Theories (In 3 Volumes)\/},  (2022), pp. 1114--1121.

\bibitem{peebles2003cosmological}
P.~J.~E. Peebles and B.~Ratra, {\em Reviews of modern physics} {\bf 75},   559  (2003).

\bibitem{armendariz2001essentials}
C.~Armendariz-Picon, V.~Mukhanov and P.~J. Steinhardt, {\em Physical Review D} {\bf 63},   103510  (2001).

\bibitem{khoury2004chameleon}
J.~Khoury and A.~Weltman, {\em Physical Review D} {\bf 69},   044026  (2004).

\bibitem{khoury2013chameleon}
J.~Khoury, {\em Classical and Quantum Gravity} {\bf 30},   214004  (2013).

\bibitem{Dima2021Dynamical}
A.~Dima, M.~Bezares and E.~Barausse, {\em Physical Review D}   (2021), \doi{10.1103/PhysRevD.104.084017}.

\bibitem{brax2017neutron}
P.~Brax, A.-C. Davis and R.~Jha, {\em Physical Review D} {\bf 95},   083514  (2017).

\bibitem{li2016neutron}
K.~Li, M.~Arif, D.~G. Cory, R.~Haun, B.~Heacock, M.~G. Huber, J.~Nsofini, D.~A. Pushin, P.~Saggu, D.~Sarenac {\em et~al.}, {\em Physical Review D} {\bf 93},   062001  (2016).

\bibitem{Vlachos2021Echoes}
C.~Vlachos, E.~Papantonopoulos and K.~Destounis, {\em Physical Review D}   (2021), \doi{10.1103/PhysRevD.103.044042}.

\bibitem{Qin2022Polarized}
X.~Qin, S.~Chen, Z.~Zhang and J.~Jing, {\em The Astrophysical Journal} {\bf 938}  (2022), \doi{10.3847/1538-4357/ac8f49}.

\bibitem{Cañate2021Gravitational}
P.~Cañate, J.~Sultana and D.~Kazanas, {\em Classical and Quantum Gravity} {\bf 38}  (2021), \doi{10.1088/1361-6382/abf97f}.

\bibitem{DeCesare2022Evolving}
M.~de~Cesare and R.~Oliveri, {\em Physical Review D}   (2022), \doi{10.1103/PhysRevD.106.044033}.

\bibitem{Davis2014Astrophysical}
A.~Davis, R.~Gregory, R.~Jha and J.~Muir, {\em Journal of Cosmology and Astroparticle Physics} {\bf 2014}, 033   (2014).

\bibitem{ottoni2024x}
T.~Ottoni, J.~G.~Coelho, R.~CR~de Lima, J.~P.~Pereira and J.~A.~Rueda, {\em The European Physical Journal C} {\bf 84},   1337  (2024).

\bibitem{maurya2021anisotropic}
S.~Maurya, K.~N. Singh, M.~Govender, A.~Errehymy and F.~Tello-Ortiz, {\em The European Physical Journal C} {\bf 81},   729  (2021).

\bibitem{barausse2013neutron}
E.~Barausse, C.~Palenzuela, M.~Ponce and L.~Lehner, {\em Physical Review D—Particles, Fields, Gravitation, and Cosmology} {\bf 87},   081506  (2013).

\bibitem{Herdeiro2020Spin-Induced}
C.~Herdeiro, E.~Radu, H.~O. Silva, T.~Sotiriou and N.~Yunes, {\em Physical review letters} {\bf 126 1},   011103  (2020), \doi{10.1103/PhysRevLett.126.011103}.

\bibitem{Wong2019Effective}
L.~K. Wong, A.~Davis and R.~Gregory, {\em Physical Review D}   (2019), \doi{10.1103/PhysRevD.100.024010}.

\bibitem{silva2022ghost}
H.~O. Silva, A.~Coates, F.~M. Ramazano{\u{g}}lu and T.~P. Sotiriou, {\em Physical Review D} {\bf 105},   024046  (2022).

\bibitem{koivisto2015scalar}
T.~S. Koivisto, E.~N. Saridakis and N.~Tamanini, {\em Journal of Cosmology and Astroparticle Physics} {\bf 2015},   047  (2015).

\bibitem{Krori1975}
K.~D. Krori and J.~Barua, {\em Journal of Physics A: Mathematical and General} {\bf 8},   508  (1975), \doi{10.1088/0305-4470/8/4/012}.

\bibitem{abbas2014cylindrically}
G.~Abbas, S.~Nazeer and M.~Meraj, {\em Astrophysics and Space Science} {\bf 354}, 449  (2014).

\bibitem{abbas2015strange}
G.~Abbas, S.~Qaisar and A.~Jawad, {\em Astrophysics and Space Science} {\bf 359},  ~57  (2015).

\bibitem{abbas2015anisotropicfG}
G.~Abbas, D.~Momeni, M.~Aamir~Ali, R.~Myrzakulov and S.~Qaisar, {\em Astrophysics and Space Science} {\bf 357}, 1  (2015).

\bibitem{zubair2016possible}
M.~Zubair, G.~Abbas and I.~Noureen, {\em Astrophysics and Space Science} {\bf 361},  ~8  (2016).

\bibitem{abbas2015anisotropic}
G.~Abbas, M.~Zubair and G.~Mustafa, {\em Astrophysics and Space Science} {\bf 358},  ~26  (2015).

\bibitem{wyman1981static}
M.~Wyman, {\em Physical Review D} {\bf 24},   839  (1981).

\bibitem{jetzer1992dynamical}
P.~Jetzer and D.~Scialom, {\em Physics Letters A} {\bf 169}, 12  (1992).

\bibitem{kodama1978general}
T.~Kodama, {\em Physical Review D} {\bf 18},   3529  (1978).

\bibitem{kodama1979properties}
T.~Kodama, L.~De~Oliveira and F.~d. Santos, {\em Physical Review D} {\bf 19},   3576  (1979).

\bibitem{antoniadis2013massive}
J.~Antoniadis, P.~C. Freire, N.~Wex, T.~M. Tauris, R.~S. Lynch, M.~H. Van~Kerkwijk, M.~Kramer, C.~Bassa, V.~S. Dhillon, T.~Driebe {\em et~al.}, {\em Science} {\bf 340},   1233232  (2013).

\bibitem{torii1999can}
T.~Torii, K.~Maeda and M.~Narita, {\em Physical Review D} {\bf 59},   104002  (1999).

\bibitem{nazar2023relativistic}
H.~Nazar, M.~Azam, G.~Abbas, R.~Ahmed and R.~Naeem, {\em Chinese Physics C} {\bf 47},   035109  (2023).

\bibitem{sakstein2013stellar}
J.~Sakstein, {\em Physical Review D—Particles, Fields, Gravitation, and Cosmology} {\bf 88},   124013  (2013).

\bibitem{sakstein2014detecting}
J.~Sakstein, B.~Jain and V.~Vikram, {\em International Journal of Modern Physics D} {\bf 23},   1442002  (2014).

\bibitem{sakstein2015testing}
J.~Sakstein, {\em Physical Review D} {\bf 92},   124045  (2015).

\bibitem{bowers1974anisotropic}
R.~L. Bowers and E.~Liang, {\em Astrophysical Journal, Vol. 188, p. 657 (1974)} {\bf 188},   657  (1974).

\bibitem{galeev2021anisotropic}
R.~Galeev, R.~Muharlyamov, A.~A. Starobinsky, S.~V. Sushkov and M.~S. Volkov, {\em Physical Review D} {\bf 103},   104015  (2021).

\bibitem{kennedy2018reconstructing}
J.~Kennedy, L.~Lombriser and A.~Taylor, {\em Physical Review D} {\bf 98},   044051  (2018).

\bibitem{liebling2023dynamical}
S.~L. Liebling and C.~Palenzuela, {\em Living Reviews in Relativity} {\bf 26},  ~1  (2023).

\bibitem{mazur2004gravitational}
P.~O. Mazur and E.~Mottola, {\em Proceedings of the National Academy of Sciences} {\bf 101}, 9545  (2004).

\bibitem{kulkarni2022if}
M.~Kulkarni, E.~Visbal, G.~L. Bryan and X.~Li, {\em The Astrophysical Journal Letters} {\bf 941},   L18  (2022).

\bibitem{ventagli2025neutron}
G.~Ventagli, P.~G. Fernandes, A.~Maselli, A.~Padilla and T.~P. Sotiriou, {\em Physical Review D} {\bf 111},   024001  (2025).

\bibitem{kiselev2003quintessence}
V.~V. Kiselev, {\em Classical and Quantum Gravity} {\bf 20}, 1187  (2003).

\bibitem{heydarzade2017black}
Y.~Heydarzade and F.~Darabi, {\em Physics Letters B} {\bf 771}, 365  (2017).

\bibitem{konoplya2019shadow}
R.~Konoplya, {\em Physics Letters B} {\bf 795}, 1  (2019).

\bibitem{zeng2020shadows}
X.-X. Zeng, H.-Q. Zhang and H.~Zhang, {\em The European Physical Journal C} {\bf 80}, 1  (2020).

\bibitem{abdujabbarov2017shadow}
A.~Abdujabbarov, B.~Toshmatov, Z.~Stuchl{\'\i}k and B.~Ahmedov, {\em International Journal of Modern Physics D} {\bf 26},   1750051  (2017).

\bibitem{chen2005quasinormal}
S.~Chen and J.~Jing, {\em Classical and Quantum Gravity} {\bf 22},   4651  (2005).

\bibitem{zhang2006quasinormal}
Y.~Zhang and Y.~Gui, {\em Classical and Quantum Gravity} {\bf 23},   6141  (2006).

\bibitem{thomas2012thermodynamics}
B.~B. Thomas, M.~Saleh and T.~C. Kofane, {\em General Relativity and Gravitation} {\bf 44}, 2181  (2012).

\bibitem{saleh2018thermodynamics}
M.~Saleh, B.~B. Thomas and T.~C. Kofane, {\em International Journal of Theoretical Physics} {\bf 57}, 2640  (2018).

\bibitem{toledo2019some}
J.~Toledo and V.~Bezerra, {\em The European Physical Journal C} {\bf 79},   110  (2019).

\bibitem{toledo2019reissner}
J.~d.~M. Toledo and V.~Bezerra, {\em International Journal of Modern Physics D} {\bf 28},   1950023  (2019).

\bibitem{sakti2020kerr}
M.~F. Sakti, A.~Suroso and F.~P. Zen, {\em Annals of Physics} {\bf 413},   168062  (2020).

\bibitem{santos2023kiselev}
L.~Santos, F.~da~Silva, C.~Mota, I.~Lobo and V.~Bezerra, {\em General Relativity and Gravitation} {\bf 55},  ~94  (2023).

\bibitem{saadati2021thin}
R.~Saadati and F.~Shojai, {\em Classical and Quantum Gravity} {\bf 38},   135025  (2021).

\bibitem{javed2024impact}
F.~Javed and M.~H. Alshehri, {\em Annals of Physics} {\bf 464},   169658  (2024).

\bibitem{dzhunushaliev2011chameleon}
V.~Dzhunushaliev, V.~Folomeev and D.~Singleton, {\em Physical Review D—Particles, Fields, Gravitation, and Cosmology} {\bf 84},   084025  (2011).

\bibitem{Boonserm2019Decomposition}
P.~Boonserm, T.~Ngampitipan, A.~Simpson and M.~Visser, {\em Physical Review D}   (2019), \doi{10.1103/PhysRevD.101.024022}.

\bibitem{campitelli2024neutron}
A.~Campitelli and L.~Mastrototaro, {\em arXiv preprint arXiv:2403.18752}   (2024).

\bibitem{riley2021nicer}
T.~E. Riley, A.~L. Watts, P.~S. Ray, S.~Bogdanov, S.~Guillot, S.~M. Morsink, A.~V. Bilous, Z.~Arzoumanian, D.~Choudhury, J.~S. Deneva {\em et~al.}, {\em The Astrophysical Journal Letters} {\bf 918},   L27  (2021).

\bibitem{Abbott2017GW170817:}
B.~A. et. al., {\em Physical review letters} {\bf 119 16},   161101  (2017), \doi{10.1103/PhysRevLett.119.161101}.

\bibitem{abbott2018gw170817}
B.~P. Abbott, R.~Abbott, T.~Abbott, F.~Acernese, K.~Ackley, C.~Adams, T.~Adams, P.~Addesso, R.~X. Adhikari, V.~B. Adya {\em et~al.}, {\em Physical review letters} {\bf 121},   161101  (2018).

\bibitem{capano2020stringent}
C.~D. Capano, I.~Tews, S.~M. Brown, B.~Margalit, S.~De, S.~Kumar, D.~A. Brown, B.~Krishnan and S.~Reddy, {\em Nature Astronomy} {\bf 4}, 625  (2020).

\bibitem{copeland2006dynamics}
E.~J. Copeland, M.~Sami and S.~Tsujikawa, {\em International Journal of Modern Physics D} {\bf 15}, 1753  (2006).

\bibitem{ghosh2016rotating}
S.~G. Ghosh, {\em The European Physical Journal C} {\bf 76},   222  (2016).

\bibitem{jamil2015dynamics}
M.~Jamil, S.~Hussain and B.~Majeed, {\em The European Physical Journal C} {\bf 75}, 1  (2015).

\bibitem{visser2020kiselev}
M.~Visser, {\em Classical and Quantum Gravity} {\bf 37},   045001  (2020).

\bibitem{caldwell2005limits}
R.~Caldwell and E.~V. Linder, {\em Physical review letters} {\bf 95},   141301  (2005).

\bibitem{holden2000self}
D.~J. Holden and D.~Wands, {\em Physical Review D} {\bf 61},   043506  (2000).

\bibitem{amendola2005constraints}
L.~Amendola, C.~Quercellini and E.~Giallongo, {\em Monthly Notices of the Royal Astronomical Society} {\bf 357}, 429  (2005).

\bibitem{babichev2005accretion}
E.~Babichev, V.~Dokuchaev and Y.~N. Eroshenko, {\em Journal of Experimental and Theoretical Physics} {\bf 100}, 528  (2005).

\bibitem{babichev2013black}
E.~O. Babichev, V.~I. Dokuchaev and Y.~N. Eroshenko, {\em Physics-Uspekhi} {\bf 56},   1155  (2013).

\bibitem{zlatev1999quintessence}
I.~Zlatev, L.~Wang and P.~J. Steinhardt, {\em Physical Review Letters} {\bf 82},   896  (1999).

\bibitem{hill1988models}
C.~T. Hill and G.~G. Ross, {\em Nuclear Physics B} {\bf 311}, 253  (1988).

\bibitem{anderson1998dark}
G.~W. Anderson and S.~M. Carroll, Dark matter with time-dependent mass, in {\em COSMO-97\/},  (World Scientific, 1998), pp. 227--229.

\bibitem{huey2000cosmological}
G.~Huey, P.~J. Steinhardt, B.~A. Ovrut and D.~Waldram, {\em Physics Letters B} {\bf 476}, 379  (2000).

\bibitem{Sagunski2017Neutron}
L.~Sagunski, J.~Zhang, M.~Johnson, L.~Lehner, M.~Sakellariadou, S.~Liebling, C.~Palenzuela and D.~Neilsen, {\em Physical Review D} {\bf 97},   064016  (2017), \doi{10.1103/PhysRevD.97.064016}.

\bibitem{Barreira2015K-mouflage}
A.~Barreira, P.~Brax, S.~Clesse, B.~Li and P.~Valageas, {\em Physical Review D} {\bf 91},   123522  (2015), \doi{10.1103/PhysRevD.91.123522}.

\bibitem{bekenstein1974exact}
J.~D. Bekenstein, {\em Annals of Physics} {\bf 82}, 535  (1974).

\bibitem{ford2001classical}
L.~H. Ford and T.~A. Roman, {\em Physical Review D} {\bf 64},   024023  (2001).

\bibitem{fewster2006averaged}
C.~J. Fewster and L.~W. Osterbrink, {\em Physical Review D—Particles, Fields, Gravitation, and Cosmology} {\bf 74},   044021  (2006).

\bibitem{jetzer1992boson}
P.~Jetzer, {\em Physics Reports} {\bf 220}, 163  (1992).

\bibitem{Collaboration2018Properties}
T.~L.~S. Collaboration, T.~V. Collaboration and B.~A. et. al., {\em Physical Review Letters} {\bf 9}, 1  (2018), \doi{10.1103/PhysRevX.9.011001}.

\bibitem{Berti2012Light}
E.~Berti, L.~Gualtieri, M.~Horbatsch and J.~Alsing, {\em Physical Review D} {\bf 85},   122005  (2012), \doi{10.1103/PhysRevD.85.122005}.

\bibitem{sennett2017distinguishing}
N.~Sennett, T.~Hinderer, J.~Steinhoff, A.~Buonanno and S.~Ossokine, {\em Physical Review D} {\bf 96},   024002  (2017).

\bibitem{Upadhye2012Quantum}
A.~Upadhye, W.~Hu and J.~Khoury, {\em Physical review letters} {\bf 109 4},   041301  (2012), \doi{10.1103/PhysRevLett.109.041301}.

\bibitem{betz2022searching}
J.~Betz, J.~Manley, E.~M. Wright, D.~Grin and S.~Singh, {\em Physical Review Letters} {\bf 129},   131302  (2022).

\bibitem{odintsov2022neutron}
S.~D. Odintsov and V.~K. Oikonomou, {\em Annals of Physics} {\bf 440},   168839  (2022).

\bibitem{glendenning2000lower}
N.~K. Glendenning, {\em Physical Review Letters} {\bf 85},   1150  (2000).

\bibitem{wiringa1988equation}
R.~B. Wiringa, V.~Fiks and A.~Fabrocini, {\em Physical Review C} {\bf 38},   1010  (1988).

\bibitem{douchin2001unified}
F.~Douchin and P.~Haensel, {\em Astronomy \& Astrophysics} {\bf 380}, 151  (2001).

\bibitem{akmal1998equation}
A.~Akmal, V.~Pandharipande and D.~Ravenhall, {\em Physical Review C} {\bf 58},   1804  (1998).

\bibitem{crawford2009neutron}
J.~P. Crawford and D.~Kazanas, {\em The Astrophysical Journal} {\bf 701},   1701  (2009).

\bibitem{read2009constraints}
J.~S. Read, B.~D. Lackey, B.~J. Owen and J.~L. Friedman, {\em Physical Review D—Particles, Fields, Gravitation, and Cosmology} {\bf 79},   124032  (2009).

\bibitem{read2009measuring}
J.~S. Read, C.~Markakis, M.~Shibata, K.~Ury{\=u}, J.~D. Creighton and J.~L. Friedman, {\em Physical Review D—Particles, Fields, Gravitation, and Cosmology} {\bf 79},   124033  (2009).

\bibitem{damour1996tensor}
T.~Damour and G.~Esposito-Farese, {\em Physical Review D} {\bf 54},   1474  (1996).

\bibitem{silva2016low}
H.~O. Silva, H.~Sotani and E.~Berti, {\em Monthly Notices of the Royal Astronomical Society} {\bf 459}, 4378  (2016).

\bibitem{chatziioannou2020neutron}
K.~Chatziioannou, {\em General Relativity and Gravitation} {\bf 52},   109  (2020).

\bibitem{abbott2020gw190425}
B.~P. Abbott, R.~Abbott, T.~Abbott, S.~Abraham, F.~Acernese, K.~Ackley, C.~Adams, R.~Adhikari, V.~Adya, C.~Affeldt {\em et~al.}, {\em The Astrophysical Journal} {\bf 892},  ~L3  (2020).

\bibitem{raaijmakers2020constraining}
G.~Raaijmakers, S.~Greif, T.~Riley, T.~Hinderer, K.~Hebeler, A.~Schwenk, A.~Watts, S.~Nissanke, S.~Guillot, J.~Lattimer {\em et~al.}, {\em The Astrophysical Journal Letters} {\bf 893},   L21  (2020).

\bibitem{choudhury2024nicer}
D.~Choudhury, T.~Salmi, S.~Vinciguerra, T.~E. Riley, Y.~Kini, A.~L. Watts, B.~Dorsman, S.~Bogdanov, S.~Guillot, P.~S. Ray {\em et~al.}, {\em The Astrophysical Journal Letters} {\bf 971},   L20  (2024).

\bibitem{mustafa2020physically}
G.~Mustafa, M.~F. Shamir and X.~Tie-Cheng, {\em Physical review D} {\bf 101},   104013  (2020).

\bibitem{horvat2013dark}
D.~Horvat and A.~Marunovi{\'c}, {\em Classical and quantum gravity} {\bf 30},   145006  (2013).

\bibitem{maurya2018generalized}
S.~Maurya, S.~Ray, S.~Ghosh, S.~Manna and T.~Smitha, {\em Annals of Physics} {\bf 395}, 152  (2018).

\bibitem{thirukkanesh2012exact}
S.~Thirukkanesh and F.~Ragel, {\em Pramana} {\bf 78}, 687  (2012).

\bibitem{noureen2019models}
I.~Noureen, S.~Mardan, M.~Azam, W.~Shahzad and S.~Khalid, {\em The European Physical Journal C} {\bf 79}, 1  (2019).

\bibitem{mardan2018new}
S.~Mardan, I.~Noureen, M.~Azam, M.~Rehman and M.~Hussan, {\em The European Physical Journal C} {\bf 78}, 1  (2018).

\bibitem{singh2020anisotropic}
K.~N. Singh, S.~Maurya, P.~Bhar and F.~Rahaman, {\em Physica Scripta} {\bf 95},   115301  (2020).

\bibitem{sharif2020anisotropic}
M.~Sharif and A.~Naeem, {\em International Journal of Modern Physics A} {\bf 35},   2050121  (2020).

\bibitem{morales2018charged}
E.~Morales and F.~Tello-Ortiz, {\em The European Physical Journal C} {\bf 78}, 1  (2018).

\bibitem{raposo2019anisotropic}
G.~Raposo, P.~Pani, M.~Bezares, C.~Palenzuela and V.~Cardoso, {\em Physical Review D} {\bf 99},   104072  (2019).

\bibitem{mak2003anisotropic}
M.~Mak and T.~Harko, {\em Proceedings of the Royal Society of London. Series A: Mathematical, Physical and Engineering Sciences} {\bf 459}, 393  (2003).

\bibitem{Maurya2018Anisotropic}
S.~Maurya, A.~Banerjee, M.~K. Jasim, J.~Kumar, A.~Prasad and A.~Pradhan, {\em Physical Review D}   (2018), \doi{10.1103/PhysRevD.99.044029}.

\bibitem{herrera1992cracking}
L.~Herrera, {\em Physics Letters A} {\bf 165}, 206  (1992).

\bibitem{abreu2007sound}
H.~Abreu, H.~Hern{\'a}ndez and L.~A. N{\'u}nez, {\em Classical and Quantum Gravity} {\bf 24},   4631  (2007).

\bibitem{tsizh2014distribution}
M.~Tsizh, B.~Novosyadlyj and Y.~Kulinich, {\em arXiv preprint arXiv:1412.7323}   (2014).

\bibitem{PhysRev.116.1027}
H.~A. Buchdahl, {\em Phys. Rev.} {\bf 116}, 1027 (Nov 1959).

\bibitem{andreasson2008sharp}
H.~Andr{\'e}asson, {\em Journal of Differential Equations} {\bf 245}, 2243  (2008).

\bibitem{pani2018gravitational}
P.~Pani and V.~Ferrari, {\em Classical and Quantum Gravity} {\bf 35},   15LT01  (2018).

\bibitem{shapiro2024black}
S.~L. Shapiro and S.~A. Teukolsky, {\em Black holes, white dwarfs and neutron stars: the physics of compact objects} (John Wiley \& Sons, 2024).

\bibitem{Eiroa2014Strong}
E.~F. Eiroa and C.~Sendra, {\em The European Physical Journal C} {\bf 74}, 1  (2014), \doi{10.1140/epjc/s10052-014-3171-1}.

\bibitem{Pang2018Gravitational}
X.~Pang and J.~Jia, {\em Classical and Quantum Gravity} {\bf 36}  (2018), \doi{10.1088/1361-6382/ab0512}.

\bibitem{Zhang2017Strong}
R.~Zhang, J.~Jing and S.~Chen, {\em Physical Review D} {\bf 95},   064054  (2017), \doi{10.1103/PhysRevD.95.064054}.

\bibitem{Sarkar2006Strong}
K.~Sarkar and A.~Bhadra, {\em Classical and Quantum Gravity} {\bf 23}, 6101   (2006), \doi{10.1088/0264-9381/23/22/002}.

\bibitem{tiede2022measuring}
P.~Tiede, M.~D. Johnson, D.~W. Pesce, D.~C. Palumbo, D.~O. Chang and P.~Galison, {\em Galaxies} {\bf 10},   111  (2022).

\bibitem{karim2025desi}
M.~A. Karim, J.~Aguilar, S.~Ahlen, S.~Alam, L.~Allen, C.~A. Prieto, O.~Alves, A.~Anand, U.~Andrade, E.~Armengaud {\em et~al.}, {\em arXiv preprint arXiv:2503.14738}   (2025).

\bibitem{tutusaus2023first}
I.~Tutusaus, C.~Bonvin and N.~Grimm, {\em arXiv preprint arXiv:2312.06434}   (2023).

\end{thebibliography}

\end{document}